\documentclass{aa}
\usepackage{pst-all}
\usepackage{color}
\usepackage{txfonts}
\usepackage{makecell}
\usepackage{amsmath}
\usepackage{amssymb}
\usepackage{bm}
\usepackage{graphicx}
\usepackage{array,multirow}
\usepackage{url}
\usepackage{siunitx}
\usepackage{soul}

\DeclareSIUnit\ergon{erg}
\bibpunct{(}{)}{;}{a}{}{,} 
\usepackage[pdfauthor={Barbosa et al.},
            pdftitle={A preserved  high-z compact progenitor in the  heart of NGC3311 revealed with MUSE 2D stellar population analysis}]{hyperref}
\hypersetup{colorlinks=true, linkcolor=blue, citecolor=blue, urlcolor=blue}
\usepackage[flushleft]{threeparttable}
\usepackage{acronym}
\acrodef{BCG}{brightest cluster galaxy}
\acrodef{BCGs}{brightest cluster galaxies}
\acrodef{BSF}{Bayesian spectral fitting}
\acrodef{ETGs}{early-type galaxies}
\acrodef{FOV}{field-of-view}
\acrodef{FWHM}{full width at half maximum}
\acrodef{HMC}{Hamiltonian Monte Carlo}
\acrodef{HST}{Hubble space telescope}
\acrodef{IFS}{integral field spectroscopy}
\acrodef{IMF}{initial mass function}
\acrodef{LOSVD}{line-of-sight velocity distribution}
\acrodef{MCMC}{Monte Carlo Markov chains}
\acrodef{SED}{spectral energy distribution}
\acrodef{SFH}{star formation history}
\acrodef{S/N}{signal-to-noise ratio}
\acrodef{SSP}{single stellar population}

\usepackage{layouts}

\definecolor{darkgreen}{rgb}{0.09, 0.45, 0.27}

\defcitealias{2016A&A...589A.139B}{B16}
\defcitealias{2018A&A...609A..78B}{B18}
\begin{document} 

\title{A preserved  high-$z$ compact progenitor in the  heart of NGC3311 revealed with MUSE 2D stellar population analysis\thanks{Table 3 is only available in electronic form at the CDS via anonymous ftp to cdsarc.u-strasbg.fr (130.79.128.5) or via http://cdsweb.u-strasbg.fr/cgi-bin/qcat?J/A+A/}}

\titlerunning{A preserved  high-$z$ compact progenitor in the heart of NGC 3311}

\author{C.~E. Barbosa\inst{\ref{usp}}\fnmsep\thanks{Corresponding author: {\tt carlos.barbosa@usp.br}.}\and
              C. Spiniello\inst{\ref{inaf_naples}, \ref{oxf},\ref{eso}} \and
              M. Arnaboldi\inst{\ref{eso}} \and
              L. Coccato\inst{\ref{eso}} \and
              M. Hilker\inst{\ref{eso}} \and
              T. Richtler\inst{\ref{concepcion}} 
              }

\institute{
Universidade de S\~ao Paulo, IAG, Departamento de Astronomia, Rua do Mat\~ao 1226, S\~ao Paulo, SP, Brazil\label{usp}
\and 
INAF, Osservatorio Astronomico di Capodimonte, Via Moiariello  16, 80131, Naples, Italy\label{inaf_naples}
\and
Department of Physics, University of Oxford, Denys Wilkinson Building, Keble Road, Oxford OX1 3RH, UK\label{oxf} 
\and
European Southern Observatory,  Karl-Schwarzschild-Stra\ss{}e 2, 85748, Garching, Germany\label{eso} 
\and
Departamento de Astronomia, Universidad de Concepci\'on, Concepci\'on, Chile\label{concepcion}     
}

   \date{Received October 30, 2020; Accepted February 26, 2021}

 
  \abstract
    {Massive early-type galaxies are believed to be the end result of an extended mass accretion history. The stars formed \textit{in situ} very early on in the initial phase of the assembly might have originated from an extremely intense and rapid burst of star formation. These stars may still be found within the cores of such galaxies at $z=0$, depending on their accretion and merger histories.}
   {We wish to investigate the presence of a surviving high-$z$ compact progenitor component in the brightest galaxy of the Hydra~I cluster, NGC~3311, by mapping its 2D kinematics and stellar population out to 2 effective radii. Our goal is to understand the formation of its several structural components and trace their mass assembly back in time.}
   {We combined MUSE observations, a customized and extended version of the state-of-the-art EMILES single stellar population models, and a newly developed parametric fully Bayesian framework to model the observed spectra using full-spectrum fitting.}
   {We present 2D maps and radial profiles of the stellar velocity dispersion, age, total metallicity, $\alpha$-element, sodium abundance ([Na/Fe]), and the \ac{IMF} slope. All properties have significant gradients, confirming the existence of multiple structural components, also including a ``blue spot'' characterized by younger and more metal-rich stars. We find that the component dominating the light budget of NGC~3311 within $R\lesssim 2.0$ kpc is the surviving $z=0$ analog of a high-$z$ compact core.
   This concentrated structure
   has a relatively small velocity dispersion ($\sigma_*\approx 180$ km s$^{-1}$), is very old (ages$\gtrsim 11$ Gyr), metal-rich ([Z/H]$\sim0.2$ and [Na/Fe]$\sim0.4$), and has a bottom-heavy \ac{IMF} (with slope $\Gamma_b\sim2.4$). In the outer region, instead, the line-of-sight velocity distribution becomes increasingly broad, and the stars are younger. They are also more metal and sodium poor but  are richer in $\alpha$-elements. The low-mass end of the IMF slope becomes Chabrier-like with increasing galactocentric distance.}
    {The existence of multiple structural components in NGC~3311 from photometry, kinematics, and stellar populations confirms the predictions from the two-phase formation scenario for NGC~3311, according to which a first very short, high-$z$ star-formation episode formed a compact stellar structure  in its core, which then grew in size by the extended mass assembly of relatively massive satellites. Interestingly, the outer stellar population has an overabundant [$\alpha$/Fe], most likely because NGC~3311, located at the center of the galaxy cluster, accreted stars from rapidly quenched satellites.}


   \maketitle
%

\section{Introduction}
%

Massive \ac{ETGs} play a crucial role in the cosmic structure formation and mass assembly of the Universe, because they account for more than half of its current total stellar mass \citep{2010ApJ...709..644I, 2011MNRAS.412..246V} and are responsible for most of its chemical enrichment \citep{2012ceg..book.....M}. They are also the oldest objects at each epoch, and thus can provide valuable insights into the star-formation activity that happened in the early Universe \citep[e.g.,][]{2006ARA&A..44..141R}.
Understanding the history of the assembly of the most massive galaxies throughout cosmic time is therefore crucial to constrain models of galaxy formation and evolution.

In this context, observational and theoretical studies in the last decade support a two-phase formation scenario for the formation of the most massive ETGs and their intriguing dynamical and stellar population properties \citep[e.g.][]{2009ApJ...699L.178N, 2010ApJ...725.2312O, 2012MNRAS.425.3119H, 2016MNRAS.458.2371R}. According to this scenario, 
a first intense, fast, and dissipative series of processes forms their central ``bulk'' mass (at $z>2$), generating, after star-formation quenches, a massive, passive, and very compact galaxy \citep[a so-called ``red nugget'';][]{2005ApJ...626..680D}. Subsequently, a second, more time-extended phase dominated by mergers and gas inflows is responsible for the dramatic structural evolution and size growth from $z\sim1$ to today \citep{2008ApJ...677L...5V, 2008ApJ...687L..61B, 2011ApJ...739L..44D, 2012ApJ...749..121S, 2018A&A...619A.137B}. 
These high-$z$ red nuggets can typically reach
masses similar to those of local giant elliptical galaxies, 
which indicates that a large portion of the mass is assembled during this first formation phase (\textit{in situ}). However, their sizes are only about a fifth of the size of local ETGs of similar mass \citep{2005ApJ...626..680D, 2008ApJ...677L...5V, 2018MNRAS.477.3886W}. Therefore, during a second phase, at lower redshifts, red nuggets must increase their size, accreting other systems and gas and finally building up, over billions of years, the present-day giant ETGs. The two-phase formation scenario successfully explains the stellar population gradients observed in massive ETGs \citep{Tortora+10, Li18, Forbes18,Ferreras19,Bernardi19,Santucci20} and also seems to naturally explain the recent observations claiming that the low-mass end of the stellar \ac{IMF} might not be universal, as has been assumed in the last 30 years, but might vary across different galaxies and radially within a single object, for massive ETGs \citep{2012ApJ...760...71C,2012ApJ...753L..32S,Barnabe+13,2014MNRAS.438.1483S,Spiniello+15, 2018MNRAS.478.4084S, 2018MNRAS.477.3954P, 2018MNRAS.479.2443V, 2019MNRAS.489.4090L}.

The study of the \textit{in situ} population in the local Universe can 
provide important insights into the physical mechanisms regulating the mass assembly  
in the early Universe. Relic galaxies, the local counterparts of red nuggets that survived untouched,  skipping 
the second phase of mass assembly, are therefore the ideal systems 
to understand the early phase of  
galaxy formation, but they are extremely rare \citep{2014ApJ...780L..20T, 2017MNRAS.467.1929F, 2021A&A...646A..28S}. Alternatively, the cores of massive \ac{ETGs} can provide information about early stellar formation in the Universe, considering that stars from red nuggets are expected to seed the most massive ETGs \citep[e.g.,][]{2009ApJ...697.1290B, 2009MNRAS.398..898H}. In particular, some of the red nuggets may still be preserved inside \ac{ETGs} in the form of compact progenitors (CPs) cores, depending on the assembly history and cosmic evolution, for example, whether the galaxy has undergone major mergers that have destroyed the compact structure or have only experienced minor- and mini-mergers which are more likely to conserve the spatial segregation between the accreted and \textit{in situ} components.

Recent analysis of the Illustris TNG simulations by \citet{2020arXiv200901823P} showed that the fraction of ETGs originating from a high-$z$ compact progenitor \citep[according to the definition of][]{Wellons16} is a strong function of stellar mass. In the range $10^{10.5}M_\sun\leq M_* \leq 10^{11} M_\sun,$ only 18\% of the ETGs were compact in the past, while at $M_*> 10^{11}M_\sun$ this fraction rises to 80\%. However, \citet{2020arXiv200901823P} also showed that the CP survives in the innermost region of the high-mass ETG in     only about 20\% of cases.



In this work, we perform a bi-dimensional (2D) mapping of the stellar population of NGC~3311, the \ac{BCG}\acused{BCGs} of the Hydra~I cluster (Abell 1060), including a varying \ac{IMF} slope, to investigate the presence of a surviving counterpart of the high-$z$ compact progenitor in the core of the galaxy. NGC~3311 has been extensively studied by our group, and previous results already indicated that it is a suitable candidate to host a CP core, as we summarize below.
The cD galaxy NGC~3311 is one of the nearest prototypical massive elliptical galaxies. Photometrically, it has an extended and diffuse stellar envelope on top of a high surface-brightness central component. As shown in \citet{2012A&A...545A..37A} and \citet[][hereafter B18]{2018A&A...609A..78B}, the surface brightness profile of NGC~3311 requires more than one component to describe its overall asymmetry with respect to the galaxy’s luminous center. 
Moreover, long-slit and \ac{IFS} observations indicate that NGC~3311  (i) has a velocity dispersion profile that rises from a central value of $\sigma_0 \sim 175$ kms$^{-1}$ to 400 kms$^{-1}$ at major axis distances of 20 kpc 
\citep{2011A&A...528A..24V,2011A&A...531A.119R}; (ii) features in the velocity dispersion profile that are correlated with photometric \citep{2015IAUS..309..221H, 2018A&A...619A..70H} and stellar population substructures (\citealt{2011A&A...533A.138C}; \citealt[][hereafter B16]{2016A&A...589A.139B}); and  (iii) that the outer halo of NGC~3311 has a peculiar line-of-sight velocity, with a measured velocity bias of $0.3$ for this BCG. These studies indicate that NGC~3311 can be described as a central galaxy surrounded by a cD envelope, similarly to IC~1101 \citep{1979ApJ...231..659D} and NGC~6166 \citep{2015ApJ...807...56B}, with the outer envelope additionally displaced from the cluster's center because of an ongoing subcluster merging in the Hydra~I core  (\citealt{2011A&A...528A..24V}, B18).

In the current project, we wish to establish connections among the photometric and kinematical  components from a stellar population point of view. 
For this purpose, we use a customized and improved version of the state-of-the-art stellar population models \citep[the eMIUSCAT, ][]{2016MNRAS.463.3409V} and 
deploy a novel method to model composite stellar populations using a full spectral fitting approach under a Bayesian framework. 

This paper is structured as follows. In \S\ref{sec:data}, we describe the observations and the data processing steps necessary for the analysis of the stellar populations of NGC~3311. 
In \S\ref{sec:sspmodels} we provide details of the \ac{SSP} models we use, while in \S\ref{sec:modeling} we introduce and describe the Bayesian method to constrain the stellar population parameters from optical spectra and \ac{SSP} models. We also explain the degeneracies between the parameters of the fitting. 
In \S\ref{sec:results}, we present the spatial distribution of the stellar populations of NGC~3311, while in \S\ref{sec:formation_scenario} we highlight the implications that they might have on the formation scenario of NGC~3311.  Finally, in \S\ref{sec:conclusion} we summarize our work.  

For the remainder of this work, we assume that NGC~3311 is at a distance of 50.7 Mpc from the Sun, based on the radial velocity of \SI{3777}{km.s^{-1}} \citep{1999ApJS..125...35S} and assuming a standard $\Lambda$CDM cosmology with $H_0=70.5$ \si{km.s^{-1}.Mpc^{-1}} \citep{2009ApJS..180..330K}. We note that while performing spectroscopic stellar population analysis, we do not exactly measure the IMF, because in old systems, only stars with masses $\sim 1 M_\odot$ are present, while more massive ones are already dead.  Hence, even though throughout the paper we always refer to the \ac{IMF} slope (or low-mass end of the \ac{IMF}), we caution the reader that we are measuring instead a composite present-day mass function.

\section{Observations and data reduction}
\label{sec:data}

We use \ac{IFS} data of NGC~3311 observed with the MUSE spectrograph \citep{2003SPIE.4841.1096H, 2004SPIE.5492.1145B} mounted on the \SI{8}{m} UT4 telescope of the Very Large Telescope and obtained under the ESO programme 094.B-0711A (PI: M. Arnaboldi). The complete observation strategy for NGC~3311 consisted of a mosaic of four pointings in wide field mode \citepalias[see details in][]{2018A&A...609A..78B}, but in this work we restrict the analysis to the two pointings with high \ac{S/N}, the central field (I) and a second pointing (II) located northeast from the center, which covers NGC~3311 out to a projected radius $R\approx 2 R_e$, assuming an effective radius of $R_e=$\SI{34}{\arcsec} \citep{2012A&A...545A..37A}. There are two exposures for each pointing, and the total exposure times are \SI{1460}{s} and \SI{1340}{s} for fields I and II, respectively.  Additionally, there are two offset pointings for sky subtraction. 

We take advantage of a new version of the individual datacubes, directly retrieved from the ESO Phase 3 archive\footnote{\url{http://archive.eso.org/cms.html}}, which were reduced with an updated MUSE pipeline \citep[v.1.4b1][]{2012SPIE.8451E..0BW, 2020A&A...641A..28W}. The data reduction performs all the general data reduction steps including bias subtraction, flat-field, sky subtraction, and wavelength and flux calibration, with uncertainties propagated throughout the pipeline. 

The reduced data cubes for NGC 3311 contain strong sky residuals at wavelengths $\lambda\gtrsim 7000$ \r{A}, which are clearly observed in the outer regions of the observations, but still noticeable even in the highest \ac{S/N} spaxels at the center of the galaxy. We clean the strongest lines with the \textsc{zap} code \citep{2016MNRAS.458.3210S} using the offset sky observations, but even after this process, over-subtracted OH line groups still dominate the spectra at longer wavelengths. However, the redder part of the optical MUSE spectra is crucial in determining variations in the low-mass end of the \ac{IMF}, as dwarf stars emit strongly at redder wavelengths \citep{1994ApJS...95..107W,2012ApJ...747...69C}. We therefore perform a second-order sky subtraction in parallel to the full spectral fitting modeling of the science data, as discussed in more detail in \S\ref{sec:modeling}. We then proceed to combine the two data cubes using customized Python and IDL scripts and the \textsc{Fits\_tools} package\footnote{\url{https://fits-tools.readthedocs.io}}. The seeing during the observations in the combined datacubes was \SI{1.08}{\arcsec} and \SI{1.26}{\arcsec} for fields I and II, respectively. 

We focus our analysis on the diffuse stellar halo of NGC 3311, but the whole area surrounding NGC 3311 contains a vast number of sources, including dwarf galaxies \citep{2008A&A...486..697M}, ultra-compact dwarfs, and globular clusters \citep{2011A&A...531A...4M}. To avoid the contamination of these systems, we have masked all compact sources detected with \textsc{sextractor} \citep{1996A&AS..117..393B} over an unsharpened-mask image of the combined data cube. 

The analysis of the \ac{IMF} variations in unresolved galaxies requires detailed modeling of absorption-line features with an accuracy of the order of 1\% \citep{2012ApJ...760...70V}. To achieve the necessary \ac{S/N} in the central pointing I, we combine data from adjacent spaxels using a Voronoi tessellation scheme with the implementation of \citet{2003MNRAS.342..345C} aiming at a \ac{S/N} of 250 per wavelength pixel for each Voronoi bin. 
This Voronoi scheme  produced a set of $74$ spectra on the central pointing.  
However, this kind of tessellation is not appropriate in pointing II, given that the stellar halo becomes very faint, and even the combination of the entire field does not produce a spectrum with a sufficiently high \ac{S/N} for our purposes. Nevertheless, to be able to cover as large  a region as possible, we still use this pointing, dividing the field of view into two sections, only considering the region within the surface brightness elliptical contour of $\mu_V=22$ arcsec$^{-2}$, which is divided in two elliptical bins. Therefore, the final dataset used in this work consists of 76 spectra.

Furthermore, the presence of a dust lane and a blue spot, a small region with relatively young stars \citep{2003AJ....125..478L, 2020arXiv200810662R}, has been reported in the center of NGC~3311.  Hence, in this region, we build up the Voronoi tessellation in a way that separates both structures (each contained in a single Voronoi bin) from the remainder of the stellar core in order to avoid contamination in the other combined spectra.

\begin{figure}
\centering
\includegraphics[width=0.99\linewidth]{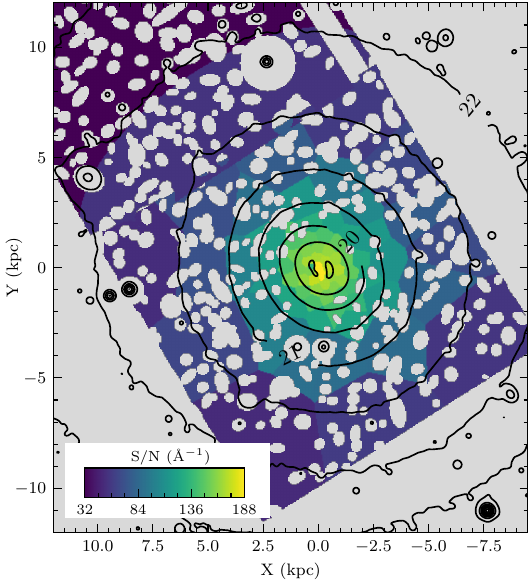}
\caption{Map of S/N for the combined spectra according to the Voronoi tessellation. Black lines indicate the $V$-band surface brightness contours from 19.5 to 22.5 mag arcsec$^{2}$ in intervals of 0.5 mag arcsec$^{2}$ from \citet{2012A&A...545A..37A}.}
\label{fig:snr}
\end{figure}

Figure~\ref{fig:snr} shows the \ac{S/N} obtained after the combination process, measured with a dispersion of 1 \r{A} with the \textsc{der\_snr} algorithm \citep{2007STECF..42....4S} and using the full wavelength range of the spectra ($4800 \lesssim \lambda (\AA) \lesssim 9400$).  If we consider only the optical regime ($\lambda (\AA) \lesssim 7000$) 
to avoid the strong sky residuals, the \ac{S/N} increases by $\sim 15$ \r{A}$^{-1}$ in the center and by $\sim 20$ \r{A}$^{-1}$ in the outskirts of the galaxy. As expected, the actual \ac{S/N} after the combination is smaller than the target value owing to the correlated noise in the data cube \citep[see][]{2013A&A...549A..87H}. 

The MUSE spectrograph has a varying spectral resolution, with a resolving power that increases from $R\approx1650$ at $\lambda=4800$ \r{A} to $R\approx3450$ at $\lambda\approx9300$ \r{A}, according to its user manual, and independently confirmed by some studies \citep[e.g.,][]{2015MNRAS.452....2K}. This translates into a spectral resolution with \ac{FWHM} in the total wavelength range between 2.5 and 3.1 \r{A}. 
To perform our analysis, we use wavelengths starting at $\lambda \approx 4800$ \r{A} which has a corresponding resolution \ac{FWHM}$\approx 2.95$ \r{A}. We homogenized the spectral resolution to this value, assuming simple Gaussian distributions for the line spread function. 
Before the analysis, we also perform two additional steps. First, considering that we model the kinematics of the spectra together with the stellar population analysis, we rebin all spectra to a logarithmic dispersion with a constant velocity scale of \SI{200}{kms^{-1}} per pixel.  This allows  
fast convolutions with the \ac{LOSVD}, similarly to the method deployed by \textsc{pPXF} \citep{2017MNRAS.466..798C}. According to previous analysis from \citet{2014ApJ...792...95C}, this rebinning process introduces small biases even in spectra binned to velocity scales of $\sim$\SI{1000}{km.s^{-1}} per pixel\footnote{We use here the notation given in the ppxf code documentation to refer to the velocity shift of each pixel.}, and therefore we do not expect to lose a significant amount of spectral information in the process.  We use the \textsc{spectres} code \citep{2017arXiv170505165C} to perform the rebinning, as it also provides a formal error propagation of the spectra. Second, we normalize all the spectra using their median flux to simplify the calculations, as this keeps the shape of the spectra unchanged, and makes the constant multiplicative factor not relevant for our analysis. 

\section{Stellar population models}
\label{sec:sspmodels}

Stellar population synthesis models are a critical ingredient to constrain the stellar population parameters: by comparing the strength of absorption features in galaxy spectra with predictions from \ac{SSP} models it is, in fact, possible to derive luminosity-weighted age, metallicity, IMF slope, and elemental abundances from a galaxy's integrated spectrum \citep{1984ApJ...287..586B, 1994ApJS...95..107W, 1998ApJS..116....1T,
2000AJ....119.1645T, 
2000AJ....120..165T, 
2000MNRAS.315..184K,
2008MNRAS.386..715T,
2010MNRAS.408...97K,
2012ApJ...760...70V, 
2012ApJ...760...71C,
2013MNRAS.433.3017L, 2014MNRAS.438.1483S, 2015ApJ...803...87S, 2016MNRAS.463.3409V,
2017ApJ...841...68V, 2017MNRAS.464.3597L}.  

In the wavelength range covered by MUSE, there are several stellar absorption features sensitive to the stellar surface gravity, that is, they are strong in dwarfs and giants and very weak in main sequence stars \citep{2014MNRAS.438.1483S}, and thus they allow us to measure the contribution of low- to high-mass stars in the integrated spectrum of a composed evolved stellar population. However, especially in the optical, the majority of these lines are also sensitive to variations in elemental abundances, metallicity, and in some cases also to age. 
Moreover, convincing evidence has emerged in recent years showing that massive ETGs can have nonsolar sodium abundance and that a high degeneracy exists between variation in the low-mass end of the IMF slope and in variation in the [Na/Fe] abundance (e.g., \citealt{2017MNRAS.464.3597L, 2018MNRAS.478.4084S}).  
A careful analysis based on a large  nondegenerate set of indicators and flexible models covering a large parameter space is therefore necessary to break the degeneracy between the stellar population parameters (age, total metallicity, [$\alpha$/Fe], and the low-mass end of the \ac{IMF} slope).  
It is also very important to have multiple tracers of the same element because different indices might show different variation with the IMF or elemental abundances, thus allowing us to break the degeneracy (e.g., the NaD is more sensitive to the sodium abundance, {[Na/Fe]}, rather than gravity; \citealt{2014MNRAS.438.1483S}).

Currently, the public version of most state-of-the-art stellar population models is not always flexible enough to allow the simultaneous variation of all the stellar population parameters. For instance, the set of models used to demonstrate for the first time the nonuniversality of the low-mass end of the \ac{IMF} \citep{2010Natur.468..940V}, the so-called CvD12 models \citep{2012ApJ...747...69C}, do not allow one to vary the total metallicity ([Z/H]) of the stellar population\footnote{However, the most recent updated version of these models presented in \cite{2018ApJ...854..139C} allows the user to vary the [Fe/H] abundance, but is not publicly available.}. On the other hand, another set of models particularly suitable for studying massive early-type galaxies, which have also been used in the past to constrain the \ac{IMF}, the MIUSCAT models \citep{2012MNRAS.424..157V, 2015MNRAS.449.1177V}, do not investigate the spectral variations due to changes in individual elemental abundances and especially in sodium, which is of crucial relevance given the Na-IMF degeneracy and the finding that massive ETGs can have Na-enhanced stellar populations.  
To overcome these limitations, and be able to simultaneously constrain age, total metallicity, $\alpha$-enhancement, [Na/Fe] abundance, and the \ac{IMF}, we combine these two sets of models, building our own custom version of the E-MILES models \citep{2016MNRAS.463.3409V} to which we add the response to sodium that we computed from CvD12, following the approach described in \citet{2015ApJ...803...87S}. We note that we assume that the spectral variation due to a change in [Na/Fe] is independent of the spectral variation due to changes in other stellar population parameters, which is not necessarily true.  

In our analysis, we use models with a double power-law ---or  ``bimodal'' \citep{1996ApJS..106..307V}--- IMF , the most versatile IMF parametrization available for the E-MILES models. The bimodal IMF is obtained by changing the high-mass end ($M \gtrsim 0.6M_{\odot}$) power-law slope ($\Gamma_b$) of the IMF with a tapered-off distribution for low-mass stars. This \ac{IMF} is virtually identical to the \citet{2001MNRAS.322..231K} IMF observed in the Milky Way for a slope with $\Gamma_b =1.3$. 
As already stressed in the introduction, performing spectroscopic stellar population analysis, we do not have constraining power on the initial conditions (i.e., the \textit{initial} mass function), because in old systems only stars with masses $\sim 1 M_\odot$ are present, while more massive ones are already dead \citep{2016MNRAS.463.3220L}.  
With the parametrization we employ here, we measure the ratio between stars with masses above and below $0.6M_\odot$. 
Thus, despite the variation formally occurring in the high-mass end of the \ac{IMF}, the change in slope is actually interpreted as a change in the fraction of stars in the low-mass range because the \ac{IMF} is normalized to $1M_\odot$.

Table \ref{tab:ssp_params} shows all the possible values for the ages, [Z/H], [$\alpha$/Fe], [Na/Fe], and the $\Gamma_b$ slope of the  IMF in our models, resulting in more than 27 000 \ac{SSP} templates. 
The original E-MILES stellar population models are produced assuming a low-mass cut-off of $0.1$M$_\odot$ and a high-mass cut-off of $100$M$_\odot$. We do not change these parameters. 

\begin{table}
\caption{Parameters of the templates used in the stellar population fitting.}
\label{tab:ssp_params}
\centering
\begin{tabular}{cp{4cm}c}
\hline
\hline
Parameter (unit)& values & \#\\
\hline
Age (Gyr)& 1, 1.5, 2.0, 2.5, 3.0, 3.5, 4.0. 4.5, 5.0, 5.5, 6.0, 6.5, 7.0, 7.5, 8.0, 8.5, 9.0, 9.5, 10.0, 10.5, 11.0, 11.5, 12.0, 12.5, 13.0, 13.5, 14.0 & 28\\

[Z/H] & -0.96, -0.66, -0.35, -0.25, 0.06, 0.15,  0.26,  0.4 & 7\\

[$\alpha$/Fe] & 0.0, 0.2, 0.4 & 3 \\

[Na/Fe] & 0.0, 0.2, 0.4 & 3 \\

$\Gamma_b$ & 0.3, 0.5, 0.8, 1.0, 1.3, 1.5, 1.8, 2.0, 2.3, 2.5, 2.8, 3.0, 3.3, 3.5 & 14 \\

Total & & 27216\\
\hline
\hline
\end{tabular}
\end{table}

\section{Bayesian full spectral fitting}
\label{sec:modeling}
 
The task of simultaneously constraining the stellar population parameters, including the IMF slope, from integrated galaxy light is a very difficult process that requires high \ac{S/N} spectra and careful stellar population modeling. Moreover, there are several degeneracies, particularly in the optical, between the IMF and other stellar population parameters \citep{2011ApJ...729..148W, 2014MNRAS.438.1483S,2015AAS...22530905T,2016ApJ...821...39M,2017MNRAS.464.3597L, 2020ApJ...902...12F}. In this context, a Bayesian full-spectrum fitting offers an appropriate method to properly determine stellar population parameters with realistic uncertainties \citep[e.g.,][]{2012ApJ...760...70V,2014ApJ...780...33C, 2017ApJ...841...68V, 2018MNRAS.479.2443V, 2018ApJ...854..139C, 2019MNRAS.485.5256Z}.

A considerable amount of information related to the stellar population and \ac{IMF} of galaxies is contained in the absorption features of the spectra. Therefore, a lot of effort has been devoted to modeling the stellar populations of \ac{ETGs} from integrated or spatially resolved spectra. Historically, many works 
have used absorption-line strengths  (in general: \citealt{1984ApJ...287..586B, 2000MNRAS.315..184K,  2000AJ....119.1645T,
2000AJ....120..165T, 2008MNRAS.386..715T, 2006ARA&A..44..141R, 2010MNRAS.408...97K}; and more recently also for the IMF in particular:  \citealt{2014MNRAS.438.1483S, 2016MNRAS.457.1468L, 2017MNRAS.465..192Z, 2018MNRAS.477.3954P, 2018MNRAS.478.4084S, 2019MNRAS.489.4090L}), also known as Lick indices, which drastically reduce the dimensionality of the modeling by focusing the analysis on a set of narrow-band features. However, more detailed modeling of the observed spectra can be performed using the full-spectral-fitting method, which consists of modeling the whole spectra\footnote{A few recent works have also adopted a hybrid approach to fitting the spectra but only at wavelengths of absorption features, e.g., \citet{2019A&A...626A.124M, 2019ApJ...875..151M}.}. This is now possible thanks to the improved quality ---in terms of flux calibration and final \ac{S/N}--- of the spectra taken with new-generation spectrographs. Even though both Lick indices and full-spectral-fitting methods may be able to recover similar results for some parameters \citep{2014ApJ...780...33C}, it is clear that the later is more able to appropriately describe the observations, as it can overcome some of the problems involved in line-strength indices, such as overlapping passbands and difficulties involving modeling in galaxies with emission lines that can contaminate the indices. Hence, full-spectrum fitting has been widely applied to most observational studies in recent years, whenever high-S/N, flux-calibrated spectra are available, and in particular to large \ac{IFS} campaigns such as SAURON \citep{2001MNRAS.326...23B}, Atlas3D \citep{2011MNRAS.413..813C}, CALIFA \citep{2012A&A...538A...8S}, SAMI \citep{2015MNRAS.447.2857B}, and MaNGA \citep{2015ApJ...798....7B},
as it provides more flexible and detailed modeling of the galaxy spectra, taking advantage of all the information available in every single spectral element.

Nevertheless, there are a few challenges involved in the full spectral fitting that have precluded its more widespread use for stellar population analyses. One important aspect is that allowing more free parameters in the model rapidly increases the number of stellar population templates required for the analysis when adopting state-of-the-art tools such as \textsc{Starlight} \citep{2005MNRAS.358..363C} or \textsc{pPXF} \citep{2017MNRAS.466..798C}, making modeling of even a single spectrum very time-consuming.
Moreover, uncertainties in the stellar population parameters are not directly available for these codes, and the error budget relies on  
extensive simulations \citep[e.g.,][]{2014A&A...561A.130C}. Consequently, only a few works have used full-spectral fitting to study the \ac{IMF} of galaxies from optical spectra so far \citep[e.g.,][]{2012ApJ...760...71C, 2017ApJ...841...68V, 2018MNRAS.479.2443V}, and none of these use the E-MILES models as we do here. Therefore, to model the observations, we take advantage of a new code, \textsc{paintbox}\footnote{\url{https://paintbox.readthedocs.io}} (Barbosa, in prep.), which is designed to overcome most of the above problems, and allows a parametric full fitting of the data using Bayesian methods. In the following sections, we explain how we proceed to perform the modeling of the observed spectra, including a detailed description of the assumptions of the model in \S\ref{sec:parametric_model}, the adopted priors and likelihood distributions used in the Bayesian analysis in \S\ref{sec:priors}, and an explanation of how we perform the modeling in \S\ref{sec:sampling}.

\subsection{Parametric model}
\label{sec:parametric_model}

We start by detailing the parametric model constructed to study the stellar population of  NGC~3311. This object is a typical passive galaxy with no indication of current star formation in most of its extension (\citealt{2011A&A...533A.138C}, \citetalias{2016A&A...589A.139B}), and therefore an \ac{SSP} approximation can be applied to describe its stellar continuum. However, there are several additional features in the spectra that should be taken into account for an accurate fitting of the observations. The central region of NGC~3311 has a dust lane with moderate emission lines and low star formation activity \citep[see][]{1991A&A...247..335V, 1994AJ....108..102G, 2009MNRAS.398..133L, 2020arXiv200810662R}. Also, as discussed in \S\ref{sec:data}, there are over-subtracted sky residuals in the redder part of the spectra, which are particularly strong in the outskirts of the field of view. Moreover, there are small differences in the flux calibration between the observations and the models that should be accounted for in a proper fitting of the absorption lines. Therefore, we model the observed galaxy spectra as

\begin{eqnarray}
 f_\lambda (\lambda) &=& \biggl (\mathcal{L}(\bm{\varv}_*) \ast S(\lambda, \bm{\theta} ) 10^{-0.4 A_\lambda (A_V, R_V)} \biggr ) \sum_{n=0}^{N} \varw_n P_n(\lambda) \\ \nonumber
 &+& \biggl [\mathcal{L}(\bm{\varv}_{\rm gas}) \ast \sum_{k=1}^K  \alpha_k G(\lambda, \lambda_k) \biggr ] + \sum_{m=1}^M \gamma_m T_m (\lambda)\mbox{.}
 \label{eq:parametric_model}
\end{eqnarray}

\noindent The first line refers to the stellar continuum, containing a \ac{LOSVD} function $\mathcal{L}(\bm{\varv_*})$ with two moments, $\bm{\varv_*}=(\varv_*, \sigma_*)$, to describe the recession velocity and the velocity dispersion; a single stellar population model $S(\lambda, \bm{\theta})$, which is parametrized by the parameters $\bm{\theta}=(\text{Age}, \text{[Z/H]}, \text{[$\alpha$/Fe]}, \text{[Na/Fe]}, \Gamma_b)$; a dust screen model to account for the internal extinction of the galaxy, assuming a \citet{1989ApJ...345..245C} extinction law $A_\lambda$, which depends on the total absorption $A_V$ and the total-to-selective extinction ratio $R_V$; and a polynomial correction of order $N$ for the flux, which is parametrized as a sum of Legendre polynomials $P_n$ with weights $\varw_n$, with $n=0,...,N$. The second line indicates the additional features necessary to properly model the continuum of the galaxy. These include  
a set of Gaussian functions, $G(\lambda_k)$, centered on emission lines and with central wavelength $\lambda_k$ and widths matching the observed resolution, which are all convolved with a second \ac{LOSVD}, 
$\mathcal{L}(\bm{\varv}_{\rm gas})$, which also has two moments, $\bm{\varv}_{\rm gas}=(\varv_{\rm gas}, \sigma_{\rm gas})$. 
In particular, we consider 
seven emission lines, including H$\beta$, H$\alpha$, [SII]6716, [SII]6731, [OIII]6300d, [OI6300]d, and [NII]6583d. Finally, the second term in the second line describes a set of templates for the sky, $T_m$, which are rescaled according to the parameters $\gamma_m$, intended to correct for the over-subtraction performed by the data reduction pipeline. 

\begin{figure}
\centering
\includegraphics{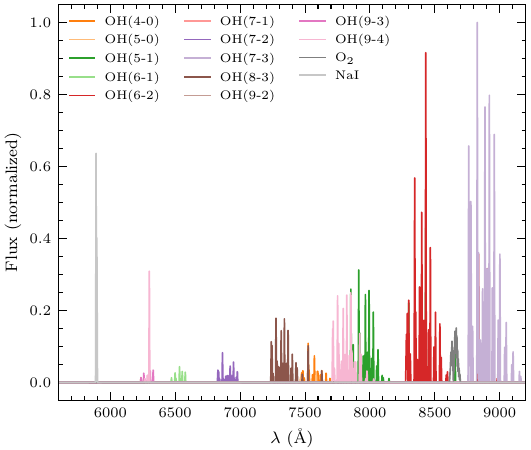}
\caption{Sky templates used in the fitting process.}
\label{fig:sky_templates}
\end{figure}

\citet{2009MNRAS.398..561W} showed that accurate sky subtraction can be achieved by simultaneous fitting with a \ac{SED} model. Our science spectra show sky-subtraction residuals at wavelengths $\lambda \gtrsim 7000$ \r{A}, most noticeably OH line groups such as OH(9-4), OH(6-2), OH(5-1), OH(9-4), and OH(8-3), but also the 0$_2$ band and the NaI doublet. To properly match the amplitude of each group of lines, we scale the amplitude of each line group independently \citep[see][]{2007MNRAS.375.1099D}.  However, separating the contribution of each individual OH group
in the observed offset sky spectrum is problematic, if one considers that 
these line groups overlap in the wavelength range, and also that they normally blend because of the limited spectral resolution and the large density of lines at some wavelengths. Thus, we instead produce a high-resolution model sky,  considering date and pointing of  our observations using the SkyCalc webtool\footnote{\url{https://www.eso.org/observing/etc/bin/gen/form?INS.MODE=swspectr+INS.NAME=SKYCALC}}, based on the Cerro Paranal Sky Model \citep{2012A&A...543A..92N, 2013A&A...560A..91J}. This model is very similar to the one obtained from the offset observations, but has the added benefit of not including noise and having precise wavelength calibration. We then identify the lines for each OH line group, the O$_2$ group, and the NaI lines using the emission line atlas from \citet{1996PASP..108..277O}. This process results in 14 sky templates that are considered in our fitting, which are plotted in Figure \ref{fig:sky_templates}. Additional sky line regions identified from sky emission atlases of \citet{1996PASP..108..277O} and \citet{2003A&A...407.1157H} are masked out of the fitting process. 

The main difference between our model and those of previous works using Bayesian full-spectrum fitting, is that we do not focus solely on the absorption features of the galaxy but, instead, we attempt to also describe the continuum of the galaxy using both the stellar continuum and an extinction law. 

The motivation for such parametrization is to increase the relevance of the shape of the continuum in the fitting process, as it also contains relevant information about the stellar population, in particular in the redder optical region and the near-infrared \citep{2018MNRAS.473.4698B}. By including an extinction law in the modeling, the role of the polynomial becomes to match the continuum locally to compensate for small differences (of the order of a few percent) between models and observations, in particular issues in the flux calibration. Moreover, by including the polynomial in the fitting, we can obtain uncertainties on the stellar population parameters marginalized over the distribution of the polynomial terms. 

As indicated by \citet{2012ApJ...747...69C}, a polynomial of order $n \sim \Delta \lambda / 100$, where $\Delta \lambda$ is the wavelength range of the spectra in Angstrom, is enough to provide a good continuum match on scales of \si{100}{\r{A},}  which is required for the analysis of absorption lines. This would imply a polynomial of order $n\sim 45$ in our case. However, after testing the modeling on a few spectra, we noticed that a polynomial of order $n=30$ provides similar residuals with a large gain in processing time, and therefore we adopt this value throughout our analysis.

\subsection{Prior and likelihood distributions}
\label{sec:priors}

In practice, we use the Bayes' theorem to determine the conditional probability of a set of parameters $\phi$ in order to describe a set of observational data $D$ using the relation:

\begin{equation}
 p(\bm{\phi} | D) \propto p(\bm{\phi}) p(D | \bm{\phi})
 \label{eq:bayes}
,\end{equation}

\noindent where $p(\phi)$ is the prior distribution, i.e., the knowledge of the distribution of parameters before the data, and $p(D|\bm{\phi})$ reflects the likelihood of the data $D$ conditional on the parameters $\bm{\phi}$. In our case, $D$ is simply the observed spectrum of a given Voronoi bin, and $\bm{\phi}$ are the parameters of the model whose priors have to be specified.

Table \ref{tab:priors} shows the prior distributions for all parameters in our model. Stellar population parameters use noninformative, uniform priors within the bounds of the models in Table \ref{tab:ssp_params}, while other stellar and gas parameters are set with very loose prior distributions based on the previous analysis of the data \citepalias{2018A&A...609A..78B}. 
For the multiplicative polynomial, we assume more restrictive priors such that the final polynomial term only slightly offsets the model to match the continuum. 

\begin{table*}
\centering
\renewcommand{\arraystretch}{1.2}
\caption{Prior distribution for all parameters in the full spectral fitting. Fluxes are normalized to the median value of the observations.} \label{tab:priors}
\begin{tabular}{ccc}
\hline
\hline
Parameter (Unit)& Prior distribution & Description\\
\hline
Age (Gyr) & $\rm{Uniform}(1, 14)$ & Age\\

[Z/H] & $\rm{Uniform}(-0.96, 0.4)$ & Total metallicity\\

[$\alpha$/Fe] & $\rm{Uniform}(0, 0.4)$ & $\alpha$-elements abundance\\

[Na/Fe] & $\rm{Uniform}(0, 0.6)$ & Sodium abundance\\ 

$\Gamma_b$ & $\rm{Uniform}(0.3, 3.5)$ & Low-mass IMF slope\\ 

$A_{\rm V}$ (mag) & $\rm{Exponential}(0.4)$ & Total extinction\\

$R_{\rm V}$ & $\rm{Normal}^+(3.1, 1)$ &Total-to-selective extinction\\

$\varv_*$ (km/s)& $\rm{Normal}(3800, 100)$ & Stellar velocity recession\\

$\sigma_*$ (km/s)& $\rm{Uniform}(100, 500)$ & Stellar velocity dispersion\\

$\alpha_k$ (\si{erg.s^{-1}.cm^{-2}.\AA^{-1}}) & $\rm{HalfNormal}(1)$ & Amplitude of the emission lines ($k=1,...,7$) \\

$\varv_{\text{gas}}$ (km/s)& $\rm{Normal}(3800, 100)$ & Gas velocity recession\\

$\sigma_{\text{gas}}$ (km/s)& $\rm{Uniform}(60, 120)$ & Gas velocity dispersion\\

$\varw_0$ (\si{erg.s^{-1}.cm^{-2}.\AA^{-1}})& $\rm{Normal}(1, 0.1)$ & Zero order polynomial weight \\

$\varw_n$ & $\rm{Normal}(0, 0.01)$ & Polynomial weight ($n=1,...,30$)\\

$\gamma_m$ (\si{erg.s^{-1}.cm^{-2}.\AA^{-1}}) & $\rm{Normal}(0, 1)$ & Amplitude for sky templates ($m=1,...,14$) \\

$\beta$ & $\rm{Normal}(0, 1)$ & Exponent of multiplicative factor for uncertainties \\

\hline \hline
\end{tabular}
\end{table*}

As we mentioned in \S\ref{sec:data}, the uncertainties obtained for the combined spectra are underestimated because they do not take into account the correlated errors between spaxels, and therefore the final \ac{S/N} is smaller than the target value before the combination. One way to deal with this problem is to rescale the uncertainties by a scalar factor \citep{2012ApJ...760...71C}. Thus, similarly to \citet{2018MNRAS.479.2443V}, we use a log-likelihood as a normal distribution with rescaled uncertainties given by 

\begin{eqnarray}
 \ln p(D | \bm{\phi}) &=& -\frac{N \ln (2\pi)}{2}  - \frac{1}{2}\sum_{\lambda=1}^N \frac{\biggl ( f_\lambda^{\text{obs}}(\lambda)-f_\lambda(\lambda, \bm{\phi})\biggr )^2}{(1+e^\beta)\,\sigma_\lambda^2} \\ \nonumber
 &-& \frac{1}{2}\sum_{\lambda=1}^N\ln \biggl ( (1+e^\beta)\,\sigma_\lambda^2 \biggr ) \mbox{,}
\end{eqnarray}

\noindent where $s^2_\lambda=(1+e^\beta)\,\sigma_\lambda^2$ is simply the uncertainty obtained from error propagation multiplied by a factor larger than one. The scale factor $\beta$ in the log-likelihood becomes a nuisance variable, whose prior distribution is also indicated in Table \ref{tab:priors}. Accounting for this additional nuisance parameter in the likelihood, the parametric model used to describe the data contains 64 free parameters.

\subsection{Model fitting}
\label{sec:sampling}

Previous works on Bayesian full-spectrum fitting were based on tools specifically developed around CvD12 and CvD16 models, such as \textsc{pystaff} \citep{2018MNRAS.479.2443V}. However, as we are using a different set of stellar population models, we use the newly developed code \textsc{paintbox} (parametric fitting toolbox, Barbosa, in prep.), an extended and more general version of the \ac{SED} fitting code used in \citet{2020ApJS..247...46B}. The \textsc{paintbox} is a simple, flexible framework for building and modeling the SEDs of galaxies, and it is intended to be used  in (though
is not restricted to) applications using Bayesian methods. It can be used to build a large range of custom parametric models for data, including multiple stellar populations, emission lines, sky or telluric templates, extinction laws, and polynomial corrections, and can be adapted to use any stellar population models.

For the \ac{MCMC} sampling of the posterior, we use the \textsc{pymc3} package \citep{Salvatier2016} to build a model with the priors detailed in Table~\ref{tab:priors}. We tested different \ac{MCMC} samplers available in the \textsc{pymc3}, including the Metropolis-Hastings \citep[MH,][]{doi:10.1063/1.1699114, 10.1093/biomet/57.1.97} algorithm and sequential Monte Carlo \citep{doi:10.1061/(ASCE)0733-9399(2007)133:7(816), 10.1093/gji/ggt180}, and also  the Afinne Invariant MCMC Emsemble sampler \citep{2010CAMCS...5...65G}, using the \textsc{emcee} code \citep{2013PASP..125..306F}. After testing different configurations for the sampling, we noticed that satisfactory convergence in a reasonable execution time is obtained by initializing the sampling with the Metropolis-Hastings algorithm, followed by a sampling with the Emsemble sampler. Both the SMC and the Affine Invariant samplers perform similarly concerning the estimation of parameters, but the Emsemble sampler has been shown to converge faster in practice, and therefore we adopt this latter throughout the analysis. 

\begin{figure*}
\centering
\includegraphics{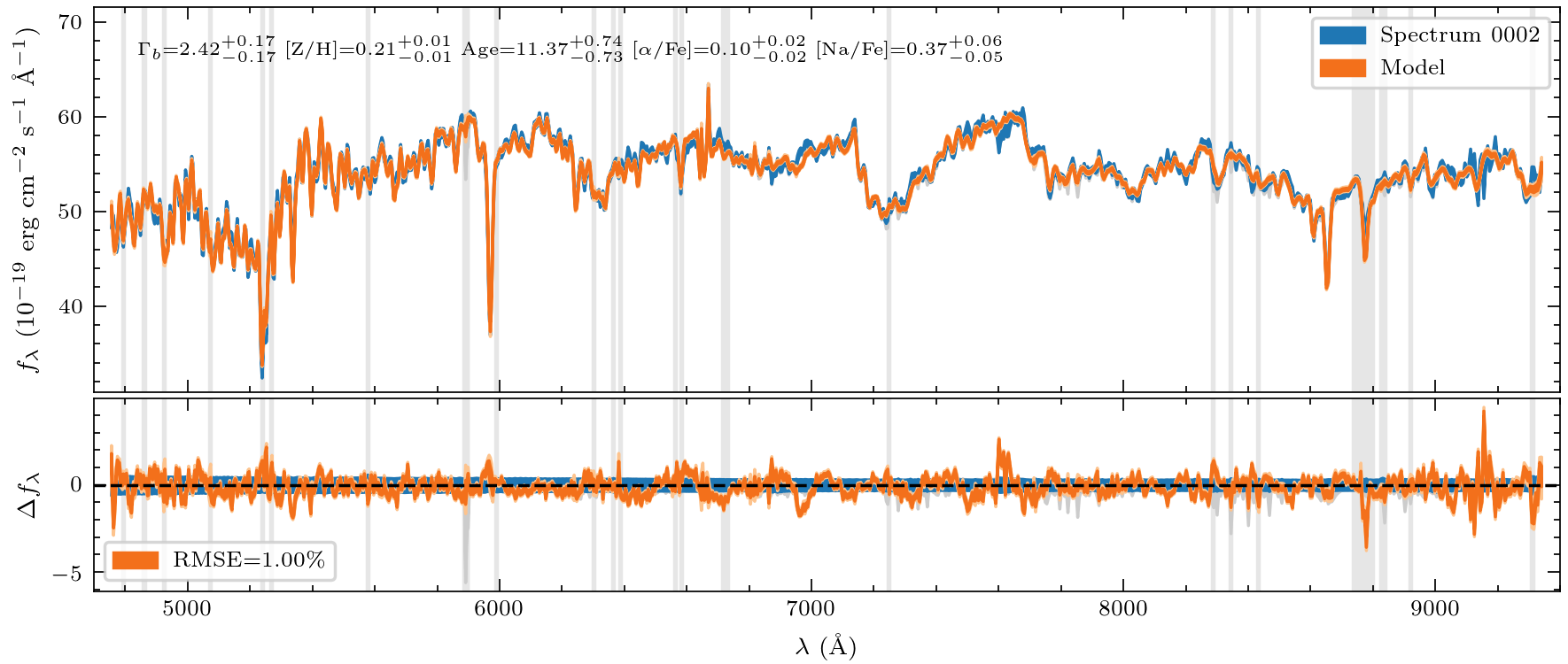}

\includegraphics{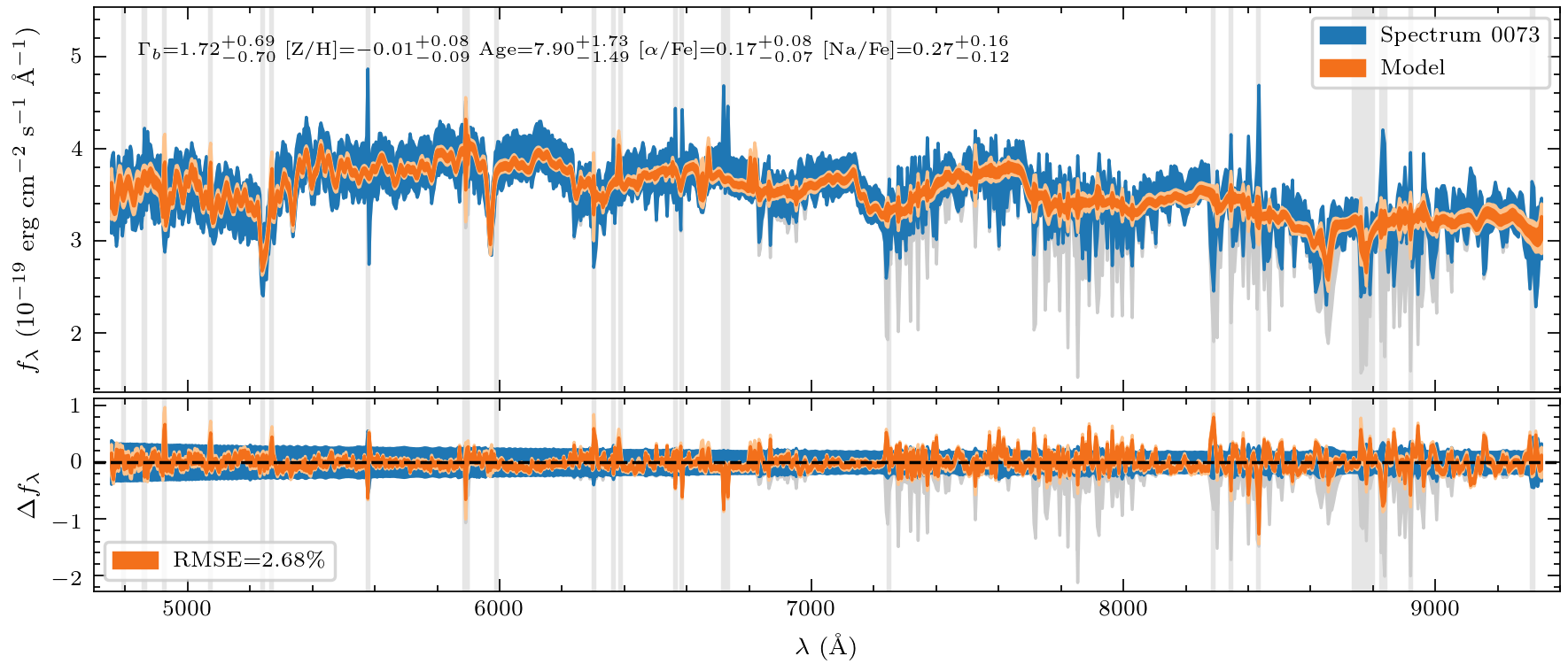}
\caption{Comparison between the observed spectrum and the best-fit model for two spectra, spectrum 0002 (upper panels) in the inner core of NGC~3311 ($R=0.4$ kpc), and spectrum 0073 (bottom panels) outside of the core ($R=8.2$ kpc) in a region where the galactic halo dominates the light. For each spectrum, the larger top panels show the observed data (dark gray), the observed data with the improved sky subtraction (blue), the best-fit models (orange), and masked sky lines (light gray vertical bands). The bottom panels display the residual uncertainty of the fitting, with the percentual root mean squared errors (RMSE) indicated in the bottom left inset. Uncertainties ($\pm 1 \sigma$) in the spectra and models are indicated by the thickness of the lines. The stellar population results are indicated at the top of each plot.}
\label{fig:fitting}
\end{figure*}

The results of our work are summarized in Table 3, distributed in electronic form. We adopt the median of the marginalized posterior distributions as estimates of the parameters, and we use the percentiles of 16\% and 84\% to represent the $\pm 1\sigma$ uncertainties. The results of the fitting for two cases are presented in Figure~\ref{fig:fitting}, one for the innermost part of the galaxy (corresponding to the Voronoi bin 0002), at an average radius of $R=0.4$ kpc, with high \ac{S/N} and emission lines, and one for the outer region in the field A (Voronoi bin 0073), at a radius of $R=8.2$ kpc, which has the strongest telluric residuals and the smallest \ac{S/N} for that field. We do not show the innermost spectrum (0001) here because this is the only spectrum severely affected by dust attenuation (see Fig.~\ref{fig:Avmap} in the Appendix), and thus it is not representative of the core of the galaxy, as we show later in \S\ref{sec:results}. 

\stepcounter{table}

In each panel, the original galaxy spectrum is plotted in blue and the best-fit template is overplotted in orange. Grey lines in the lower panel show the telluric residuals, which were removed during the modeling. The comparison between the model and the observations indicates that the fitting can reproduce the observed data well, with residuals of only a few percent over almost the entire wavelength range of the spectra in both cases, as indicated by the root mean squared errors (RSME). The RSME, shown in the top right inset of each panel, is estimated using the median absolute deviation, which is robust against outliers in the distribution. We stress that Spectrum 0073 is the worst case we have and we show it in Figure~\ref{fig:fitting} to highlight the effect of the over-subtracted telluric contamination in the outer region of field I. The typical residuals for the fitting are usually more similar to Spectrum 0002. For the whole sample, the median RSME of the residuals is 1.4\%, and RSME larger than 2\% are only found in the outermost nine spectra.

\subsection{Degeneracies in the model fitting}
\label{sec:degeneracies}
Besides the determination of a best-fit model and realistic errors for all parameters, the mapping of the posterior distribution allows the verification of correlations between parameters which, in the domain of stellar populations, have been known as degeneracies, such as the age--metallicity degeneracy \citep{1994ApJS...95..107W}. Interestingly, despite the use of Bayesian tools in several previous works involving the \ac{IMF}, there have been no attempts to quantify these degeneracies, which, as we discuss in more detail in a companion letter \citep{2021A&A...645L...1B}, lead to some confusion about the main factors driving the \ac{IMF} in \ac{ETGs}. Below, we describe how we use the posterior distribution in our fitting to evaluate the effects of degeneracies in our analysis. 

\begin{figure*}
\centering
\includegraphics[width=0.49\linewidth]{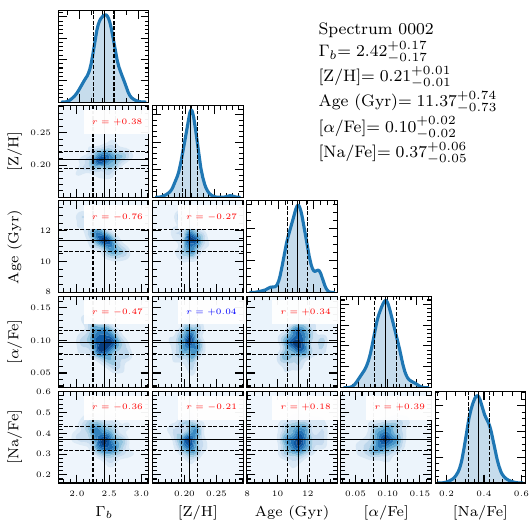}
\includegraphics[width=0.49\linewidth]{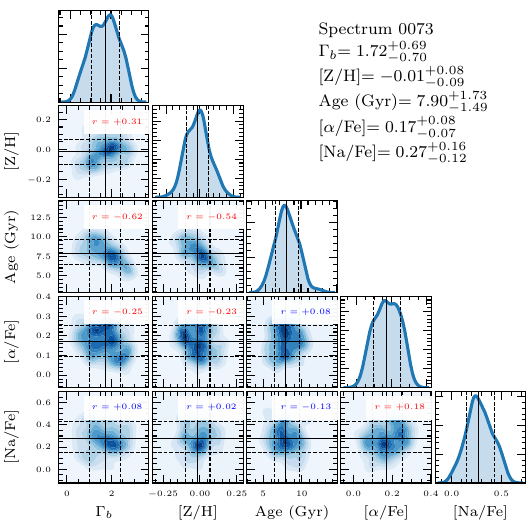}
\caption{Examples of corner plots for the same two spectra as in Figure~\ref{fig:fitting}. The diagonal histograms indicate the marginalized posterior samples used for inference, while the density maps below them represent projections of the marginalized posterior distribution. Solid lines indicate the median value used as a reference, and the dashed lines indicate the \num{16}\% and \num{84}\% percentiles used to estimate the $\pm 1\sigma$ uncertainties. The Pearson correlation coefficient $r$ is indicated at the top of each panel under the diagonal, with red (blue) colors indicating a $p$-value smaller (larger) than 1\%. A summary of the statistics for the stellar population parameters is displayed in the upper right part of each panel.}
\label{fig:corner}
\end{figure*}

Figure~\ref{fig:corner} shows the marginalized posterior distribution of the stellar population parameters, obtained for the same two spectra presented in \S\ref{sec:sampling}. The histograms on the diagonal indicate that the marginal posterior of each stellar population parameter, including the \ac{IMF} slope, are unimodal and favor models well within the ranges of the stellar population parameter limits.  However, the marginalized posterior over pairs of stellar parameters, indicated in the panels under the diagonal, clearly show the presence of several correlations among parameters, even in the high-S/N cases (i.e., Spectrum 0002, left panel), in particular involving the age, metallicity, and the IMF slope.  To measure these correlations, we report in each plot the Pearson correlation coefficient ($r$), which is often used in statistical analysis to measure the linear correlation between two variables. The coefficients of each panel are color-coded in red when they indicate a p-value smaller than 1\% to indicate which correlations are less likely to occur by chance, and in blue for larger p-values that could occur by chance. 

The correlations between parameters indicated in Figure~\ref{fig:corner} are consistent for most of the spectra in our sample, even for spectra with different stellar population properties. In Table \ref{tab:correlations}, we show the median correlation between pairs of parameters, ranked according to their statistical significance. We also include the velocity dispersion in the table to illustrate how it is mostly uncorrelated to all stellar population parameters. Interestingly, the strongest correlation is found between the age and the IMF slope, which are anti-correlated, with a correlation coefficient larger than that of the age--metallicity degeneracy, which is much more well known but ranks second in our study. 

To describe the correlation between parameters, we performed an ellipse fitting to the marginalized posterior distribution for all pairs of stellar population parameters, $\theta_1$ and $\theta_2$, measuring the semi-major ($a$) and the semi-minor ($b$) axis lengths, and the angle ($\omega$) between the semi-major axis direction about $\theta_1$, for each spectrum. The mean and the standard deviation of the ellipse parameters for the whole sample are also presented in Table \ref{tab:correlations}. We return to the consequences of these correlations in \citet{2021A&A...645L...1B}, while we proceed here with the interpretation of the results on all the stellar population parameters and their implications for the star formation history and mass assembly of NGC~3311.

\begin{table*}
\caption{Correlation between parameters in the posterior distribution. Representative values are the median of all spectra in the sample, and uncertainties are the standard deviation of all measurements. (1-2) Pair of stellar population parameters considered. (3) Pearson's $r$ correlation coefficient. (4) $p$-value relative to the correlation. (5) Semi-major axis length ($1\sigma$) of the covariance ellipse. (6) Semi-minor axis length ($1\sigma$) of the covariance ellipse. (7) Angle of the semi-major axis in relation to $\theta_1$.} \label{tab:correlations}
\centering
\renewcommand{\arraystretch}{1.2}
 \begin{tabular}{ccccccc}
 \hline
 \hline 
 $\theta_1$ & $\theta_2$ & $r$ & $p$ & $a$ & $b$ & $\omega$ ($^{\rm o}$) \\
 (1) & (2) & (3) & (4) & (5) & (6) & (7) \\
 \hline
$\Gamma_b$ & Age (Gyr) & $-0.63\pm0.15$ & $0.00\pm0.00$ & $3.25\pm0.88$ & $0.56\pm0.38$ & $-80.22\pm3.60$\\

[Z/H] & Age (Gyr) & $-0.50\pm0.22$ & $0.00\pm0.05$ & $3.21\pm0.86$ & $0.08\pm0.06$ & $-89.05\pm0.60$\\

$\Gamma_b$ & [Z/H] & $0.35\pm0.24$ & $0.00\pm0.10$ & $0.76\pm0.45$ & $0.08\pm0.07$ & $2.61\pm1.92$\\

[Na/Fe] & [$\alpha$/Fe] & $0.30\pm0.13$ & $0.00\pm0.15$ & $0.21\pm0.09$ & $0.07\pm0.05$ & $7.44\pm3.46$\\

$\Gamma_b$ & [$\alpha$/Fe] & $-0.22\pm0.16$ & $0.00\pm0.20$ & $0.75\pm0.45$ & $0.07\pm0.05$ & $-1.48\pm1.06$\\

$\Gamma_b$ & [Na/Fe] & $-0.21\pm0.15$ & $0.00\pm0.19$ & $0.76\pm0.45$ & $0.20\pm0.09$ & $-3.51\pm2.57$\\

[Na/Fe] & [Z/H] & $-0.15\pm0.16$ & $0.01\pm0.26$ & $0.21\pm0.09$ & $0.09\pm0.07$ & $-4.45\pm11.72$\\

Age (Gyr) & [Na/Fe] & $0.07\pm0.16$ & $0.07\pm0.26$ & $3.21\pm0.85$ & $0.21\pm0.09$ & $0.29\pm0.65$\\

$\sigma_*$ (km/s) & [Na/Fe] & $0.05\pm0.13$ & $0.12\pm0.29$ & $20.54\pm39.83$ & $0.21\pm0.09$ & $0.02\pm0.08$\\

Age (Gyr) & [$\alpha$/Fe] & $-0.04\pm0.18$ & $0.02\pm0.24$ & $3.21\pm0.85$ & $0.07\pm0.05$ & $-0.06\pm0.30$\\

$\Gamma_b$ & $\sigma_*$ (km/s) & $-0.04\pm0.14$ & $0.11\pm0.31$ & $20.54\pm39.83$ & $0.75\pm0.44$ & $-89.94\pm0.29$\\

[Z/H] & $\sigma_*$ (km/s) & $0.04\pm0.15$ & $0.08\pm0.30$ & $20.54\pm39.83$ & $0.09\pm0.07$ & $-90.01\pm0.05$\\

$\sigma_*$ (km/s) & Age (Gyr) & $0.03\pm0.15$ & $0.11\pm0.30$ & $20.55\pm39.82$ & $3.11\pm0.86$ & $0.10\pm1.59$\\

$\sigma_*$ (km/s) & [$\alpha$/Fe] & $0.03\pm0.10$ & $0.29\pm0.31$ & $20.54\pm39.83$ & $0.07\pm0.05$ & $0.01\pm0.02$\\

[$\alpha$/Fe] & [Z/H] & $-0.02\pm0.20$ & $0.01\pm0.29$ & $0.10\pm0.07$ & $0.07\pm0.05$ & $-73.51\pm35.18$\\

\hline 
\hline
 \end{tabular}
\end{table*}

\section{Mapping the stellar populations of NGC 3311}
\label{sec:results}
In this section we present the spatial distribution of the stellar population properties of NGC~3311, both from 2D maps and extracting the 1D spatial profiles, and then discuss the main implications on the processes involved in its formation and mass assembly.

\subsection{Two-dimensional maps of the stellar population parameters}
\label{sec:2d_maps}
\begin{figure*}
\centering
\includegraphics[trim=0.0cm 0.80cm 0.0cm 0, clip, scale=0.91]{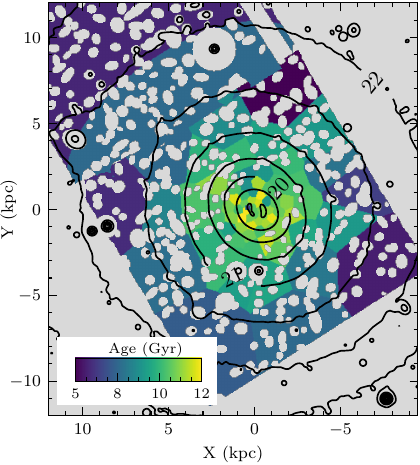}
\includegraphics[trim=0.8cm 0.80cm 0.0cm 0, clip, scale=0.91]{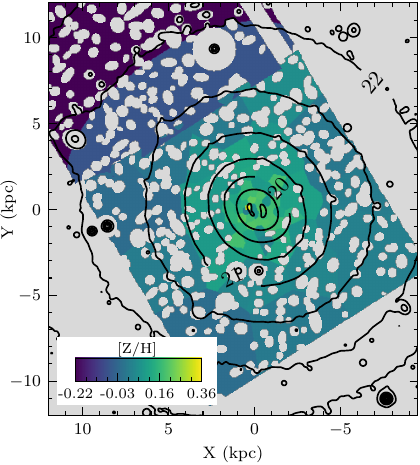}
\includegraphics[trim=0.8cm 0.80cm 0.0cm 0, clip, scale=0.91]{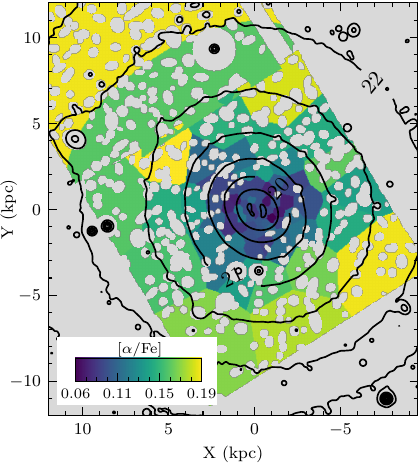}

\includegraphics[trim=0.0cm 0.00cm 0.0cm 0, clip, scale=0.91]{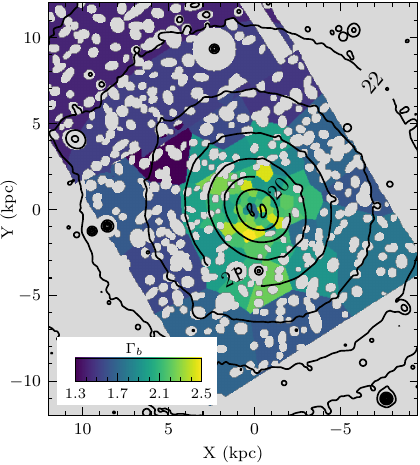}
\includegraphics[trim=0.8cm 0.00cm 0.0cm 0, clip, scale=0.91]{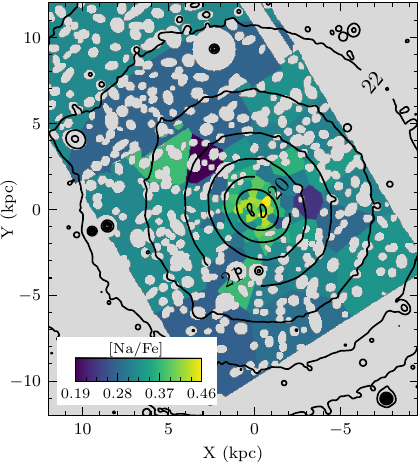}
\includegraphics[trim=0.8cm 0.00cm 0.0cm 0, clip, scale=0.91]{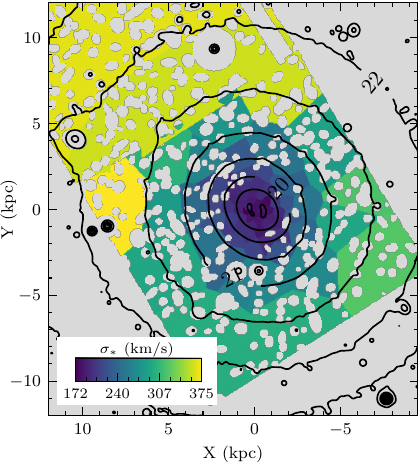}
\caption{2D maps of the stellar populations parameters in NGC~3311, including age, metallicity ([Z/H]), $\alpha$-element abundance ([$\alpha$/Fe]), low-mass IMF slope ($\Gamma_b$), sodium abundance ([Na/Fe]), and stellar velocity dispersion ($\sigma_*$).}
\label{fig:2dmaps}
\end{figure*}

The main results of our analysis are detailed 2D maps of the stellar population parameters in NGC~3311. Figure~\ref{fig:2dmaps} shows the 2D maps for five relevant stellar population parameters plus the stellar velocity dispersion map. In NGC~3311, the stellar population properties show similar behavior, with radial gradients that approximately follow the surface brightness contours, without significant azimuthal variations. 

\begin{figure*}
\centering
\includegraphics[trim=0.00cm 0.00cm 0.0cm 0, clip, scale=0.91]{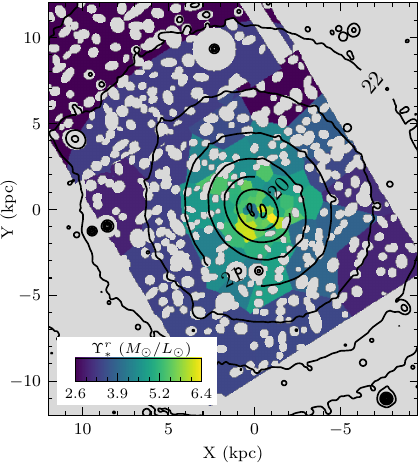}
\includegraphics[trim=0.8cm 0.00cm 0.0cm 0, clip, scale=0.91]{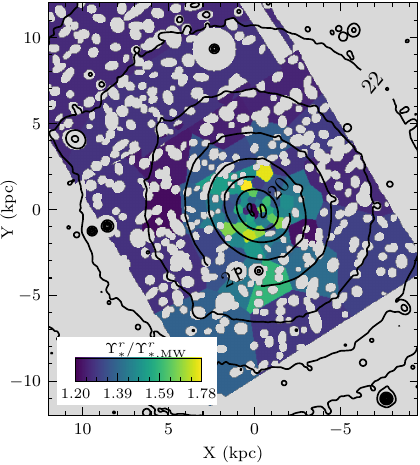}
\includegraphics[trim=0.8cm 0.00cm 0.0cm 0, clip, scale=0.91]{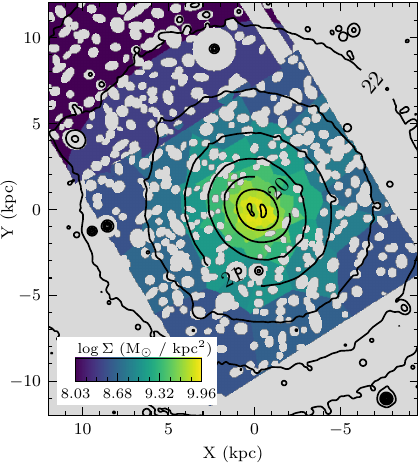}
\caption{2D maps of the mass-to-light ratio, mismatch parameter and stellar density in NGC~3311.}
\label{fig:2dmaps_2}
\end{figure*}

Interestingly, the innermost region ($R\le 0.5$ kpc) of the galaxy, where the \textit{in situ} population should dominate, shows a larger scatter with respect to bins at larger radii. 
This is likely due to the presence of the dust lane which obscures parts of this region, and to the presence of features that do not belong to the pristine \textit{in situ} population, including the  star-forming blob and the jet \citep{2003AJ....125..478L, 2020arXiv200810662R}. These stellar features indicate a relatively recent, small, star formation event which adds to the overall pristine stellar population in the center. We further speculate on this in Sect. 5.3, showing that our code is able to correctly trace the stars with intermediate age.

Similarly, the result obtained for the outermost region ($R>7$ kpc) is also characterized by a larger scatter, which we attribute both to the relatively low \ac{S/N} of the spectra, but also to the expected, stochastic contribution of accreted populations, which already dominate the light at $R\sim 1 R_e$ (see  \citetalias{2018A&A...609A..78B} and \citealt{2018A&A...619A..70H}). 

\subsection{Additional parameters derived from fitting}
\label{sec:add_param}

Besides the parameters that come directly from the fitting of the data, we also compute additional parameters of interest using the posterior distribution samples obtained from our fitting. These include the mass-to-light ratio, the mismatch parameter, and the stellar surface density, which we show in Figure~\ref{fig:2dmaps_2}. Below we explain how these quantities are derived and briefly discuss their characteristics.

The $r$-band stellar mass-to-light ratio ($\Upsilon_*^r$) shown in the left panel of Figure~\ref{fig:2dmaps_2} was obtained directly from the E-MILES SSP models \citep{2010MNRAS.404.1639V, 2012MNRAS.424..172R} as a function of the stellar population parameters in the posterior samples. In NGC~3311, $\Upsilon_*^r$ is not spatially constant but decreases with increasing galactocentric distance. It has a relatively flat value of $\sim 5.5 M_\odot / L_\odot$ in the innermost $R\sim 1.5$ kpc, while it reaches values of $\sim 3 M_\odot / L_\odot$ in the outermost region.  
Our results are consistent with the mass-to-light gradients observed in other massive \ac{ETGs} \citep[e.g.,][]{2017ApJ...841...68V, 2018MNRAS.477.3954P}.
We also note that the $\Upsilon_*^r$ values for NGC~3311 are larger than the corresponding values obtained for the stellar disks of the giant spirals NGC~628 and NGC~6946 also, where \citet{2018MNRAS.476.1909A} and \citet{2021MNRAS.500.3579A} measured surface mass densities and showed that the stellar disks are maximal. 

We then use the posterior distribution samples to determine the mass-to-light ratio of a stellar population with the same age and metallicity as inferred from the fitting but with Charier IMF ($\Upsilon_{*,\rm{Chab}}^r$). 
We therefore define a mismatch parameter, which is given by the ratio between these two mass-to-light ratios, $\Upsilon_*^r/\Upsilon_{*,\rm{MW}}^r$, to evaluate the change in mass-to-light due only to variations in the IMF low-mass end slope. The 2D map of the values of the mismatch parameter is shown in the middle panel of Figure~\ref{fig:2dmaps_2}. It is interesting to note that the mismatch parameter is always greater than one, which means that the stellar $\Upsilon_*^r$ of NGC~3311 is always larger than the value computed using a ``universal'' (Chabrier) IMF slope.

Finally, we show 
the surface stellar density for each bin  in the right panel of the same figure. This quantity is derived from the apparent and total magnitude, luminosity distance, and $\Upsilon_*^r$. The apparent $r$-band magnitude is obtained by convolving the flux-calibrated spectra of each bin with the $r$-band transmission curve using the package \textsc{speclite}\footnote{\url{https://speclite.readthedocs.io}}. The absolute magnitude of each bin is then determined by assuming a distance to NGC~3311 $D=50.4\pm10$ Mpc using the dispersion of values in the NED database to estimate the uncertainty. Finally, assuming $M_{\odot,r}=4.65$ \citep{2018ApJS..236...47W}, and the $\Upsilon_*^r$ posterior sample derived above for each bin, we obtain the stellar surface density. Like the $\Upsilon_*^r$, also this quantity has a larger value and a flat gradient in the innermost region of the galaxy ($\Sigma \sim10^{10}$ M$_{\odot}$ kpc$^{-2}$) and then shows a radial gradient towards larger radii\footnote{The surface mass density of the stellar component in NGC~3311 is 1.35 dex larger than the surface mass density values measured in the central region of NGC~6946 \citep{2021MNRAS.500.3579A} and 1.5 dex larger than the measured $\Sigma$ value in NGC~628 \citep{2018MNRAS.476.1909A}.}.

\subsection{One-dimensional spatial profiles}
\label{sec:1d_profiles}
\begin{figure}
\centering
\includegraphics[width=\columnwidth]{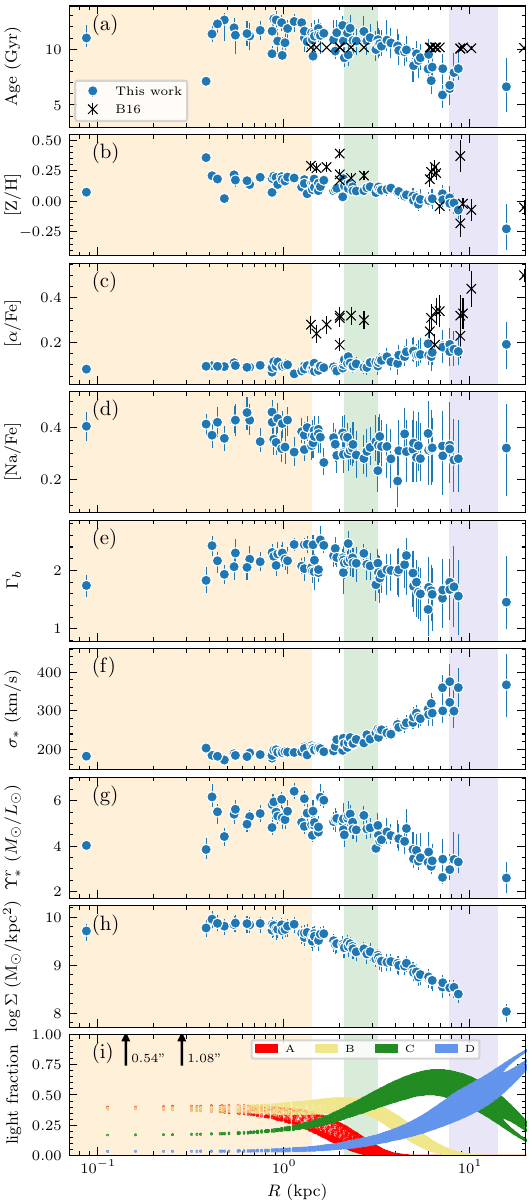}
\caption{Radial profiles of the age (a), metallicity (b), [$\alpha$/Fe] (c), [Na/Fe] (d), \ac{IMF} slope (e), velocity dispersion (f), $r$-band mass-to-light ratio (g), and stellar surface density (h) in NGC~3311. Black crosses in the three top panels show the results from previous FORS2 slitlets data \citepalias{2016A&A...589A.139B}. The bottom panel (i) shows the fraction of light for each substructure as determined from the photometric decomposition in \citetalias{2018A&A...609A..78B}. The orange shaded area indicates the region where the compact progenitor contributes significantly to the light budget, while the green and blue vertical shades indicate the transition region between dominant subcomponents B to C, and C to D. The vertical arrows in panel (i) indicate apertures with $1\times$ and $2\times$ the seeing radius (FWHM/2). We note that the x-axis is on log-scale.}
\label{fig:rprof}
\end{figure}

To better visualize the transition between the \textit{in situ} and accreted populations, also taking into account the uncertainties on the derived stellar population parameters we show, in Figure~\ref{fig:rprof}, the 1D radial profiles of the stellar population parameters (panels a to e), the stellar velocity dispersion (panel f), the $\Upsilon_*^r$ (panel g), and the surface density (panel h).  We assume that NGC~3311 is approximately round, considering that it has an ellipticity $\varepsilon\approx0.1$ in most of the field of view \citep[see][]{2012A&A...545A..37A}, and therefore for radial distances we use the projected galactocentric radii, defined as the distance of the barycenter of the Voronoi bins to the center of NGC 3311.

Moreover, we include the fraction of light of each of the four photometric components associated with NGC~3311 (A, B, C, D) as derived in \citetalias{2018A&A...609A..78B} (panel i). Components A and B have very similar surface brightness in the center, but A is more compact than B ($R_{e,A}\approx1.5$ kpc and $R_{e,B}\approx3$ kpc). In all the panels of Figure~\ref{fig:rprof}, we show an orange band to indicate the region where component A is brighter than all others ($R\lesssim 1.5$ kpc). We also indicate, with the green and purple bands in all panels, the transition between the two most dominant structures at larger radius, B to C, and C to D. In Sect.~\ref{subsec:gradients} we discuss the implications of the presence of multiple components on the mass assembly of NGC~3311.

In the central region, in particular within the region where component A dominates the light, all stellar population properties are relatively flat. As indicated by the arrows in panel (i), this flattening is not a pure effect of the PSF, but extends to $\approx 5$ FHWM of the seeing, indicating a spatially resolved core\footnote{We caution the reader that the innermost Voronoi bin (0001) at $R\approx 0.1$ is affected by dust attenuation (see Fig.~\ref{fig:Avmap}) and therefore the results for that point cannot be relied upon to understand the nature of the innermost region.}. 

\begin{figure}
\centering
\includegraphics[width=\columnwidth]{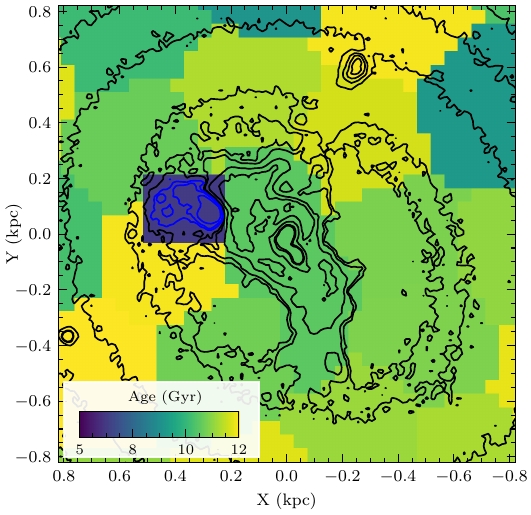}
\includegraphics[width=\columnwidth]{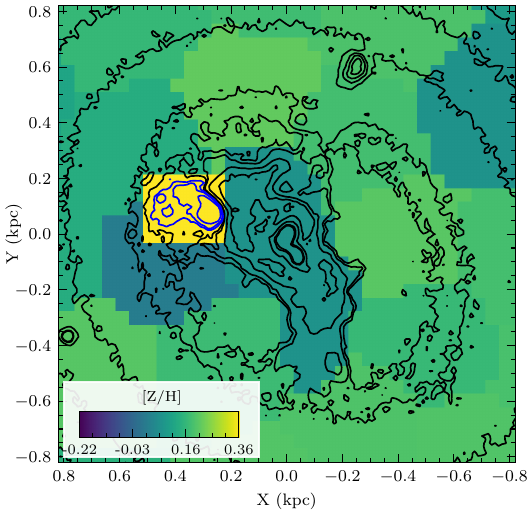}
\caption{Zoom-in of the age (top) and metallicity (bottom) 2D maps with dimension $1.6\arcsec \times 1.6 \arcsec$ including \ac{HST} contours. The blue spot \citep{2020arXiv200810662R} indicated by the blue contours is younger and more metal-rich than the surrounding regions in the core of the galaxy.}
\label{fig:age_zoom-in}
\end{figure}

In the age and metallicity 1D profiles (two upper panels of Fig.~\ref{fig:rprof}), it is possible to note one single Voronoi bin at $R\sim0.45$ kpc (within the flat innermost region) whose luminosity-weighted age (metallicity) is much younger (higher) than that of all other points up to $R\sim 3$ kpc from the center. 
This spatially corresponds to the blue spot found from {HST} images in  \citet{2020arXiv200810662R}, where there is evidence for a stellar population with intermediate age, which also spatially coincides with the region where  \citet{2005A&A...439...59H} found a fraction of intermediate-age globular clusters. This can be confirmed in  Figure~\ref{fig:age_zoom-in}, where we show a $1.6\arcsec \times 1.6 \arcsec$ zoom-in of the age (top) and metallicity (bottom) 2D map with the {HST} contours overplotted. The small region delimited by the blue continuous line highlights the blue spot, which, despite the much lower ground-based resolution, indeed corresponds to the Voronoi bin for which we find a younger ($\sim 7$ Gyr) and more metal-rich ([Z/H]$\sim 0.36$) population. The high metallicity is compatible with a scenario according to which this population comes from mass loss from asymptotic giant branch stars, and is thus enriched in metal. Accreted stars formed later on in time would instead have been more metal poor,  given the low mass of the blue spot (i.e., mass--metallicity relation). The blue spot region seems to have also a slightly bottom-light IMF, although still consistent within the error with the overall population at the same distance from the center.

Beyond $R\sim1.5$ kpc, all plotted quantities show non-negligible gradients.
In particular, stellar gradients are negative for the age, metallicity, IMF slope, sodium abundance, mass-to-light ratio, and stellar surface density, while they are positive for the $\alpha$-element abundance and the stellar velocity dispersion. 
Qualitatively, the age,  metallicity, and \ac{IMF} radial gradients are similar to those observed in several early-type galaxies from previous studies covering up to $\sim 1 R_e$ \citep[e.g.,][]{2010MNRAS.408..272S, 2015MNRAS.448.3484M, 2017ApJ...841...68V, 2018MNRAS.478.4084S, 2019MNRAS.483.3420P}. However, it is rare to have a bright central galaxy with rising $\sigma_{\star}$ and [$\alpha$/Fe] profiles with radius. \footnote{Examples of cD galaxies with rising profile known in the literature are IC~1101 \citep{1979ApJ...231..659D} and NGC~6166 \citep{2015ApJ...807...56B}.}.

For age, metallicity, and $[\alpha$/Fe], we show for comparison the results obtained in the previous analysis from FORS2 slitlets data \citepalias{2016A&A...589A.139B}. However, we stress that many differences exist between the two analyses: we use two different data-analysis techniques, namely full spectral fitting here versus Lick indices in \citetalias{2016A&A...589A.139B} and two different stellar population models, namely extended E-MILES here compared to the \citet{2011MNRAS.412.2183T} models in the previous paper. 
Up to 6 kpc, both age and metallicity results are consistent with the previous analysis. The age gradient we derive is slightly steeper for larger radii, while for the metallicity it is very hard to say anything given the large offset found between the different data points in  \citetalias{2016A&A...589A.139B}.

\begin{figure*}
 \includegraphics[width=\textwidth, trim= 0 0cm 0 0, clip]{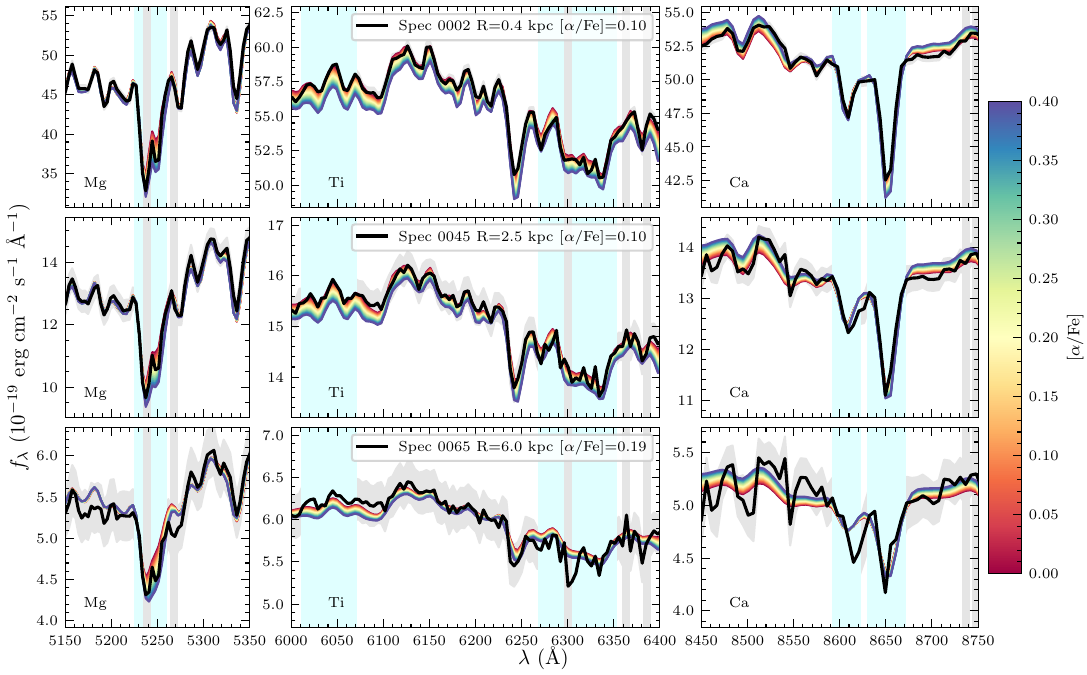}
 \caption{Zoom-in around three $\alpha$ sensitive absorption-features which are dominated by three different elements: Mg (left panels), Ti (middle panels), and Ca (right panels). Blue shaded regions highlight the bandpass of the indices often measured around these features (Mg$_b$, TiO1, TiO2, Ca1, Ca2), while gray vertical lines are regions  contaminated by sky and telluric lines and therefore  masked out from the fit. From the top to the bottom panel, we plot three NGC~3311 spectra in black, with a common resolution of 200 km/s/pixel and increasing distance from the center, as specified in the caption. Colored lines are instead MILES \ac{SSP} models with fixed age, metallicity, [Na/Fe], and IMF, and color-coded by [$\alpha$/Fe] (red solar and blue 0.4, as highlighted by the color bar on the side). The figure demonstrates that the only slightly super solar [$\alpha$/Fe] ($\sim 0.1$) inferred from the fit for the central core is a combination of a larger [Mg/Fe] plus roughly solar [Ti/Fe] and [Ca/Fe]. This explains the apparent disagreement between the [$\alpha$/Fe] obtained here via full spectral fitting and that obtained in \citetalias{2016A&A...589A.139B} using line indices, which is, in reality, a [Mg/Fe] ratio because it was computed from the Mg$b$ and Fe lines only.}
 \label{fig:alphaFe_fitting}
\end{figure*}

The most discrepant result is found for the $\alpha$-element abundance profile, with a systematic offset ($>0.1$ dex) despite the similar positive gradient. 
Such an offset can be explained considering the different techniques used in these two papers. When using line indices to constrain the [$\alpha$/Fe] value in the optical (4800-5800\AA), as done in \citetalias{2016A&A...589A.139B}, one is mainly measuring [Mg/Fe], with the largest contribution coming from the Mg$b$ line. Differently, when using full spectral fitting on a wider wavelength range, as done here, one accounts for the contribution from the continuum, the several faint molecular TiO lines, and the Ca lines, which are also $\alpha$-elements. Therefore, assuming that a stellar population may have Mg-enhanced abundance but lower [Ti/Fe] and/or [Ca/Fe] abundances, the final inferred $[\alpha$/Fe] values would be lower when determined using full spectral fitting. 
We illustrate that indeed this is the case for NGC~3311 in Figure~\ref{fig:alphaFe_fitting}, that is, stars (in the core) have an overall [$\alpha$/Fe] ratio almost consistent with solar but larger [Mg/Fe]. In the figure, we show three zoom-in panels of three representative spectra (with increasing galactocentric distance from top to bottom) around three absorption lines which are dominated by the most relevant $\alpha$ elements: the Mg$b$ ($\lambda\sim 5177$\AA), dominated by magnesium, the TiO1 and TiO2 molecular features ($\lambda\sim5940$-5995\AA\, and $\lambda\sim6190$-6415\r{A}), dominated by titanium, and the first two lines of the calcium triplet ($\lambda\sim 8500$ and $\sim8540$\r{A})\footnote{the third Ca feature is masked out (from the plot and the fit) because it is severely contaminated by telluric lines.}, clearly dominated by calcium. We note that we plot the spectra at their observed wavelengths. We overplot \ac{SSP} models with fixed age, metallicity, IMF, and [Na/Fe] (equal to those of the best-fit model for that given spectrum) and color-coded by their [$\alpha$/Fe] (red solar, blue 0.4, as highlighted by the color bar). In the core, at $R=0.4$ kpc (top panels), where we infer a  [$\alpha$/Fe]=0.1, which corresponds to the orange model, the zoom into the Mg region (left inset) requires a larger abundance, and the spectrum is better matched by the green line, which corresponds to an \ac{SSP} model with [$\alpha$/Fe]$\approx0.3$. In the central and right insets, instead, where the Ti and Ca lines are shown, the data are perfectly fit with the red line, implying a solar [$\alpha$/Fe]. The same is certainly valid also for the middle panel, showing a spectrum at $R=2.5$ kpc.  Thus, given that \citetalias{2016A&A...589A.139B} is measuring [Mg/Fe]$\sim0.3$ at that radius, we do find very similar results. For the spectra at $R=6.0$ kpc plotted in the bottom panel, it is somehow harder to reach definitive conclusions by simply looking at the plot, because the data are much noisier and contaminated by telluric lines in the redder part\footnote{these are however properly modeled in the fitting.}. However, in this case, it looks like the [Mg/Fe] might be even larger ($\sim 0.35$), which is once again in good agreement with \citetalias{2016A&A...589A.139B}. We stress that this figure is meant only as a qualitative comparison between the models and the data, and is useful to show that the quantities denoted as [$\alpha$/Fe], measured via line-indices \citepalias{2018A&A...609A..78B} and full spectral fitting (here), physically differ, and that the estimate of  [$\alpha$/Fe] from SSP models might change if obtained from different wavelength regions, where different elements are more or less dominant. In future works, it will be necessary to test \textsc{paintbox} with different SSP models which allow  the single element abundances to be changed in an independent way, which is not possible with the E-MILES ones. 
Finally, we know that in general, the fact that the [Mg/Fe] is not tracking [$\alpha$/Fe] also qualitatively agrees with the individual abundance ratio determinations by \citet{2014ApJ...780...33C}

\subsection{Multiple components in NGC~3311 from the stellar populations }
\label{subsec:gradients}

Figure~\ref{fig:rprof} clearly shows that the stellar parameters characterizing the innermost core are different from those characterizing more external regions. As stated above, all the stellar population properties are flat up to R$\sim1.5$ kpc and then they show positive or negative spatial gradients at larger radii. Moreover,  the gradients sometimes seem to have a change of slope around $R\sim3$ (e.g., for age and \ac{IMF}). This is not surprising, as the kinematic and photometric decomposition of NGC~3311 \citepalias{2018A&A...609A..78B} implies 
a superposition of multiple structural components in the central region of the galaxy to explain the measured rising velocity dispersion profile and the 2D line-of-sight velocity distribution, as shown in the last panel of Figure~\ref{fig:rprof}.

As discussed in the introduction,  the modeling of the surface brightness profile along the photometric major axis performed in  \citetalias{2018A&A...609A..78B} supports three inner components related to the central galaxy and a fourth component consistent with a cD envelope. 
Within the innermost $2.8$ kpc, two components are equally dominating the light (A, B). These have slightly different S\'ersic indices, axis ratios, and orientation, but similar kinematics ($\sigma_{A+B}\approx 153$ km s$^{-1}$). Component A is also more concentrated and small ($R\sim1.5$ kpc). At around $R=3$ kpc, a galactic halo component with velocity dispersion of $\sigma\approx 188$ km $^{-1}$ and different systemic velocity ($\Delta V = - 20$ kms$^{-1}$; named C) starts to give a large contribution to the light, as one can see from panel (i) in Figure~\ref{fig:rprof} (green curve and vertical shaded region). This component dominates up to $R_e=9.6\pm 0.3$ kpc, after which an outer stellar envelope, the cD halo (D) with $R_e=51\pm5$ kpc, a velocity dispersion of $\sigma\approx 327$ kms$^{-1}$, and $\Delta V \simeq +120$ kms$^{-1}$ takes over (see the blue curve and vertical shaded region in panel (i)).

As we required very high \ac{S/N} for the current analysis, we do not reach into the cD halo. This outer component starts to dominate the light at $R\ge 10$ kpc \citep{2018A&A...609A..78B,2018A&A...619A..70H}, where we only have one data point for which the inferred stellar population properties have considerably larger uncertainties. The large uncertainties make it impossible to recover precise information for the stellar population parameters for the outermost D component. 
We therefore cover a region where the light mainly comes from the central core and the galactic halo, as shown in the bottom panel of Figure~\ref{fig:rprof} where we reproduce the structural components from \citetalias{2018A&A...609A..78B}, and obtain the transition radii. Such radii are highlighted in all panels as vertical colored regions. The stellar population results support the presence of these structural components, with the inner core having older and slightly more metal-rich stars ---distributed following a more dwarf-rich IMF--- than the galactic halo.  

Structural transitions between inner and outer regions in massive ETGs are 
also supported by the extended kinematics measurements out to six effective radii on average from the ePN.S early-type galaxy survey \citep{2018A&A...618A..94P}, which reported the kinematic diversity of stellar halos and the presence of a transition radius in ETGs. The current analysis of the stellar populations in NGC~3311 indicates that different kinematics and structural components   also highly correlate with different stellar population parameters. We discuss them in turn below.

\subsection{The sodium abundance}
\begin{figure}
\centering
\includegraphics[width=\columnwidth]{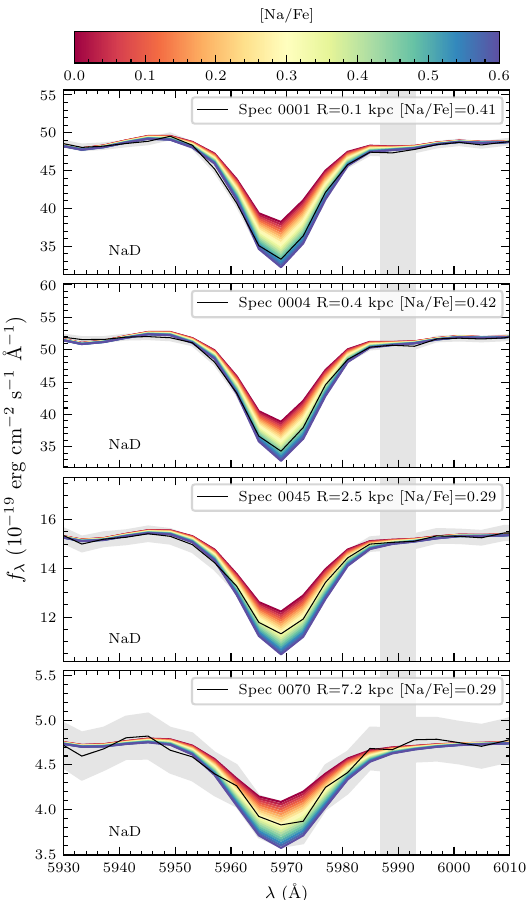}
\caption{Zoom-in around the NaD absorption line for the
spectra of four galaxies at different distances from the center (in black). Colored lines are MILES \ac{SSP} models with fixed age, metallicity, [$\alpha$/Fe], and IMF, and color-coded by [Na/Fe] (red solar and blue 0.6, as highlighted by the color bar at the top). The upper panel shows the Voronoi bin  corresponding to the central dust lane (0001). There, the contribution to the NaD line is not only stellar but also comes from the cold gas. A minimal contribution can  also be seen in the second panel (Spec 0004), but in all other bins shown, in which the dust lane is not present, the contribution is stellar with a super-solar [Na/Fe] abundance value .}
\label{fig:NaD_zoom-in}
\end{figure}

The sodium abundance in the central region of massive elliptical galaxies has recently been the topic of active discussion, not only concerning the possible [Na/Fe]--IMF slope degeneracy \citep{2015ApJ...803...87S, 2017MNRAS.464.3597L}, but also the detection of interstellar absorption, including the possible presence of cold-gas outflows \citep{2013ApJS..208....7J, 2015ApJ...809...91P, 2016MNRAS.456L..25S, 2019MNRAS.486.1608N}.  In the center of M87, \citet{2018MNRAS.478.4084S} found a very highly sodium-enhanced  stellar population, with a declining profile from [Na/Fe]$\sim 0.7$ at 2$\arcsec$ to $\sim 0.45$ at 30$\arcsec$.  While we do not find such large values in the center of NGC~3311, 
[Na/Fe] abundance is always super-solar, even at $R>R_e$
with a declining profile, although with significant scatter.  
We do not think that this can be entirely due to the central dust lane \citep{2003AJ....125..478L}, because it does not extend so far out. It is, in fact, restricted to a single Voronoi bin (0001), while the super-solar Na is found in all the spectra we analyzed, as clearly shown in Figure~\ref{fig:NaD_zoom-in}. This latter figure shows a zoom-in around the NaD line ---which shows the strongest optical sodium absorption--- of the galaxy spectra for four Voronoi bins at different distances from the center of NGC~3311. The SSP models with fixed age, metallicity, IMF, and [$\alpha$/Fe] are over-plotted and color-coded by [Na/Fe]. In particular, in the Spec~0001 (upper panel), which is the one corresponding to the dust lane, the NaD absorption is deeper than what it would be when assuming the [Na/Fe] value inferred by \textsc{paintbox}.  
While the inference from stellar population modeling is [Na/Fe]= 0.41, as indicated by the caption, the feature in the galaxy spectrum better matches a model with larger [Na/Fe] ($\sim0.55$, green). A similar behavior is observed in Spec 0004, as shown in the second panel from the top. There, the disagreement between the depth of the line and the [Na/Fe] measured from \textsc{paintbox} is smaller, because the contribution of the dust lane is minimum/null in this bin.
We therefore conclude that such high Na-enhancement is caused by the fact that NaD has  two contributions in that region: a narrower one from a gaseous disk \citep{2020arXiv200810662R}, and a broader one from the stars, characterized by different line-broadening, which sum up and thus return a better fit with models with larger [Na/Fe] abundance. 
However, we also note that the NaD line is  also very deep for the other spectra, plotted in the lower two panels, which are extracted from a region free from the dust lane. There, the [Na/Fe] values that one would infer by looking at the depth of the line perfectly fit the inference given by our stellar population modeling, that is, the NaD has only a {stellar} contribution, which is nevertheless super solar\footnote{We note that we cannot exclude the presence of a galactic diffuse component of cold atomic gas that contributes to the depth of the NaD absorption also at larger distances.}.  

\subsection{The positive [$\alpha$/Fe] gradient}

While the large majority of measured [$\alpha$/Fe] gradients are negative (i.e., $\alpha$-elements decreasing with radius) 
in massive \ac{ETGs} \citep{2013ApJ...776...64G, 2015ApJ...808...26R},   broadly consistent with predictions from an inside-out scenario, chemical evolution models predict a variety of different gradients for the [$\alpha$/Fe] and [Z/H] profiles; see \citet{2008A&A...484..679P}. The latter behavior depends on whether the inflow velocity of the gas is fast relative to the star formation. In such a case the stars that are still forming at inner radii have time to form out of $\alpha$-enhanced gas coming from the outer regions, thus flattening or even reversing the sign of the [$\alpha$/Fe] gradient \citep{2008A&A...484..679P}.

For NGC~3311, we measure a positive [$\alpha$/Fe] gradient. Ongoing star formation at the level of $1.7\times 10^{-3}$M$_\odot$yr$^{-1}$ is measured within the $2.0 \times 2.0$ kpc$^2$ central region, plus the presence of an intermediate stellar population, as already mentionned. Thus one finds stars formed \textit{in situ} but at later cosmic times from cold gas in the very center of NGC~3311. 

The cold gas traced by the NaD absorption \citep{2020arXiv200810662R} may come from the cooling of the relatively hot gas from the mass loss of late phases of stellar evolution (like AGB winds and PN envelopes; see \citealt{2015ApJ...803...77C} and \citealt{2000ASPC..199..423D}), or X-ray cooling flows \citep{2015ApJ...811..111M}.  The availability of cold gas implies that one can expect stars with younger ages and lower [$\alpha$/Fe] but not necessarily low metallicity at the very center of NGC 3311. Differently from M87, the lack of an active AGN in the galaxy center allows the hot medium to cool and give birth to new stars.
While the bulk of stars in the center of NGC~3311 may have been formed quickly in a burst of star formation in the past, the ongoing level of star formation has generated more iron and thus decreased the [$\alpha$/Fe] overabundance value there.

This result together with the larger [$\alpha$/Fe] values in the external region related to the late accretion of quenched satellites gives rise to the positive 
[$\alpha$/Fe] gradient measured in this work.

\section{Implications on the formation scenario of NGC~3311}
\label{sec:formation_scenario}

The formation of massive \ac{ETGs} entails an extended mass accretion history. At high $z$, an \textit{in situ} component is formed at their centers and from then on, further accretion and mergers shape their structure and kinematics at all radii \citep{2009ApJ...699L.178N, 2010ApJ...725.2312O}. 

Using the Illustris TNG100 cosmological simulations, \citet{2020arXiv200901823P} showed that 57\% of ETGs  are dominated by the {\it in situ} stars in the central regions, and by the {\it ex situ} stars in the outskirts.
Furthermore they showed that $\sim20\%$ of high-mass ($M_{\star} \ge 10.6 \times 10^{10}M_{\odot}$) ETGs have a so-called surviving CP in their cores, which still contains $\sim 50\%$ of the stars initially tagged as belonging to the high-$z$ CPs. 
In \citet{2020arXiv200901823P}, the high-$z$ CPs are selected on the basis of the so-called compactness parameter, defined as $\Sigma/R_{e}^{1.5}$; see also \citet{Wellons16}. High-z CPs are those for which $\log(\Sigma/R_{e}^{1.5}) > 10.5$. A surviving CP core is then a $z=0$ structure where at least 50\% of CP stars remain, leading to a compactness value of $\log(\Sigma/R_{e}^{1.5}) \simeq 10.2$.

We now turn to the structural parameters for the core of NGC~3311 which is associated with the photometric component $A+B$ in B18. The   absolute magnitude in the $V$-band of the (A+B) central component is $M_{tot} = -19.9$, and, considering an average $(V-r)$ color of $0.7$, we obtain a total luminosity in $r$-band of $L_r = 1.15 \times 10^{10}  L_{\mathrm{sun}}$.  
The effective radius for A+B is $R_e = 2.8$ kpc \citepalias[from ][]{2018A&A...609A..78B}; we also adopt the value $\Upsilon_*^r = 6$ taken from the 2D maps presented in Section~\ref{sec:add_param}. With these parameters, we infer a compactness value\footnote{using Eq.~5 \citet{2020arXiv200901823P}} for the (A+B) component in the core of NGC~3311 of 10.2, which is 
consistent with the value expected for a surviving CP component in a galaxy core.

Thus, in conclusion, 
considering the joint evidence coming from the stellar populations, kinematics, and photometry, NGC~3311 is fully consistent with the predictions of the two-phase mass assembly. First, because NGC~3311 is spatially located in the vicinity of the center of the Hydra~I cluster, the galaxy is dominated by accreted stars in its outskirts, with the velocity dispersion of its outer halo reaching a value very similar to that measured for the satellite galaxies in the subcluster core \citep{2012A&A...545A..37A}.
Secondly, the light from the very central part of the galaxy 
is dominated by a surviving CP  
that formed \textit{in situ} early on in cosmic time ($\approx 11$ Gyr ago). This structure, with $R_e\sim 3$ kpc \citepalias{2018A&A...609A..78B}, has a comparatively low velocity dispersion ($\sigma_*\approx 180$ \si{km s^{-1}}) with respect to the central $\sigma$ values of other \ac{BCGs} \citep[e.g.,][]{2009MNRAS.398..133L, 2018MNRAS.478.4084S}. 
Most likely, this compact core was formed during the first phase of the mass assembly of NGC~3311 from a short and very efficient star formation episode, as testified by the high [Mg/Fe] abundance, which however does not correspond to an equally high  $[\alpha$/Fe] abundance. 
Finally, the \ac{IMF} slope in the central part is bottom-heavy ($\Gamma_b\approx 2.4$) and consequently the stellar mass-to-light ratio is larger there ($\Upsilon_*^r \simeq 6$). 
We further investigate the \ac{IMF} radial profile and the possible physical motivation behind a nonuniversality of the IMF in massive ETGs in a companion letter \citep{2021A&A...645L...1B}.

Outside the core, our MUSE data probe mostly the galactic stellar halo \citep{2018A&A...609A..78B, 2018MNRAS.478.4255J}, which has a slightly hotter velocity dispersion ($\sigma_*\approx 190$ \si{km s^{-1}}). Here, at $R>3$ kpc, the age, metallicity, Na-abundance, and IMF slope of the stellar population decrease, while [$\alpha$/Fe] increases.  
Despite the large uncertainties on the individual data points,  from the consistency of the results we can infer that the galactic halo is a few gigayears younger than the core region ($\approx 8$ Gyr) and has roughly solar total metallicity. Interestingly, however, both the $\alpha$-element and the sodium abundances remain super solar. 
At $R\approx$ 5-6 kpc, where this component has the largest fraction of light relative to either the inner core or the cD outer halo \citepalias[see ][and panel (i) of Fig.~\ref{fig:rprof}]{2018A&A...609A..78B}, some of the stellar population properties show a flattening of their radial gradient ($\alpha$-element abundance, IMF slope, and sodium abundance). 

According to the two-phase formation scenario, the outer parts were built on top of the galactic core from extended mass accretion and mergers, whose properties shape the kinematics and structure of the resulting ETG at $z=0$ \citep{2010ApJ...725.2312O, 2009ApJ...699L.178N, 2020arXiv200901823P}. In particular, for NGC~3311, the accreted material has a larger stellar velocity dispersion, 
is more $\alpha$-enhanced, but less metal-rich than the center. As discussed in \citetalias{2016A&A...589A.139B}, the accreted material is inhomogeneous, because it contains materials from satellites with a range of masses and metallicities.  
\citet{2018A&A...619A..70H} observed a large scatter on small spatial scales in the velocity dispersion of NGC~3311 at $1R_e<R<2R_e$. These latter authors speculated that this behavior could be caused by a superposition of the core of NGC~3311 ($\sigma_{\star}\le200$ km/s$^{-1}$) with kinematical cold substructures at various velocities that mimic a high velocity dispersion in the line of sight (up to $\sigma_{\star}\sim400$ km/s$^{-1}$ at 2$R_{e}$). 
This would also explain the scatter in metallicities.  

However, it is interesting  that the velocity dispersion of the accreted material is larger than that measured at $R=0$ in the core of NGC~3311, but still significantly smaller than the velocity dispersion of the cluster itself, $\sigma_{\rm cluster}=724$ \si{km .s^{-1}} \citep{2003ApJ...591..764C}, indicating that only the closest satellites to the central core of the Hydra~I cluster contributed to the build-up of the envelope of the BGC \citep{2011A&A...528A..24V}.

\section{Summary and Conclusion}
\label{sec:conclusion}
In this paper, we present a stellar population analysis of NGC~3311, the central galaxy of the Hydra~I cluster, in which we investigate the mass assembly history of its different structural properties and verify the presence of a surviving compact progenitor using high-S/N MUSE observations. 

We performed a parametric full spectral fitting using \textsc{paintbox} (Barbosa, in prep.), adopting Bayesian methods and a customized and extended version of the publicly available simple stellar population E-MILES models. We performed our fitting of the observed spatially resolved spectra using a detailed model with 65 free parameters, including the line-of-sight velocity distribution of stars and gas, and single stellar population models with varying age, metallicity, $\alpha$-element, sodium abundances, and IMF slope (assuming a bimodal form). At the same time, we corrected for the presence of strong telluric lines in the redder part of the spectra and took into account the internal extinction of the galaxy. 

From the 2D maps and 1D radial profiles of the stellar populations and stellar velocity dispersion, we confirmed the presence of three main structural components, contributing to the light at different galactocentric distances, supporting the findings of \citetalias{2018A&A...609A..78B} from photometry and kinematics. Our findings can be summarized as follows: 
\begin{itemize}
    \item[i)] We identified a preserved compact progenitor, the surviving $z=0$ analog of the high-$z$ compact core that initially formed NGC3311, which  still dominates in the innermost $R\le1.5$ kpc, and has a very old stellar population of supersolar metallicity, Mg, and Na abundances. The low-mass end of the IMF slope of this component is  bottom-heavy and the mass-to-light ratio in $r$-band is $\approx 6 M_\odot / L_\odot$; 
    \item[ii)]  We  confirmed the presence of a  blue spot at $R\sim0.45$ kpc from the center, which has a stellar population with intermediate luminosity-weighted age (7 Gyr),  higher metallicity ([Z/H]$\sim 0.36$) and slightly bottom-light IMF ($\Gamma_b\sim1.82$) than the remaining of the core.
    \item[iii)] We identified a galactic halo, which dominates between 2.5 and 10 kpc, and is formed by relatively younger stars (7-8 Gyrs) with slightly lower metallicity and [Na/Fe] ratio (however both still supersolar) than the core but larger [$\alpha$/Fe]; 
    \item[iv)] We identified the outer halo, which dominates at radii larger than 10 kpc, whose stars are younger, more metal and sodium poor, distributed with a bottom-light IMF, but are more $\alpha$-enhanced. 
\end{itemize}

We finally discussed the implication of these results on the assembly of NGC~3311, confirming, at least for this galaxy, the two-phase scenario proposed to address the extended mass assembly of massive \ac{ETGs} \citep{2009ApJ...699L.178N, 2010ApJ...725.2312O, 2012MNRAS.425.3119H, 2016MNRAS.458.2371R, 2020arXiv200901823P}. According to our stellar population results, the innermost core was formed \textit{in situ} during a short star-formation burst, and with a bottom-heavy IMF \citep[see also the companion letter,][]{2021A&A...645L...1B}. The galactic halo component and the outer halo were formed and accreted to the outer region of NGC~3311 in a second and more time-extended phase, likely from satellites that were already quenched before the merging event with NGC~3311, as indicated by their large [$\alpha$/Fe].

\begin{acknowledgements}
The authors wish to acknowledge the anonymous referee for the constructive report. CEB gratefully acknowledges the S\~ao Paulo Research Foundation (FAPESP), grants 2011/51680-6,  2016/12331-0, 2018/24389-8. 
CS is supported by a Hintze Fellowship at the Oxford Centre for Astrophysical Surveys, which is funded through generous support from the Hintze Family Charitable Foundation.  
TR acknowledges support from the BASAL Centro de Astrofisica y Tecnologias Afines (CATA) PFB-06/2007.\\
This work is based on observations collected at the European Organisation for Astronomical Research in the Southern Hemisphere under ESO programme 094.B-0711(A). 
It has made use of the computing facilities of the Laboratory of Astroinformatics (Instituto de Astronomia, Geof\'isica e Ci\^encias Atmosf\'ericas, Departamento de Astronomia/USP, NAT/Unicsul), whose purchase was made possible by FAPESP (grant 2009/54006-4) and the INCT-A. This research has made use of the NASA/IPAC Extragalactic Database (NED), which is operated by the Jet Propulsion Laboratory, California Institute of Technology, under contract with the National Aeronautics and Space Administration.
\\
\textit{Additional software:} Astropy \citep{2013A&A...558A..33A, 2018AJ....156..123A}, matplotlib \citep{4160265}, numpy \citep{5725236}, scipy \citep{scipy}.
\end{acknowledgements}

\bibliographystyle{aa} 
\bibliography{biblio}

\appendix
\section{Additional figures}
\label{sec:app1} 
\begin{figure}
\centering
\includegraphics[width=0.94\linewidth]{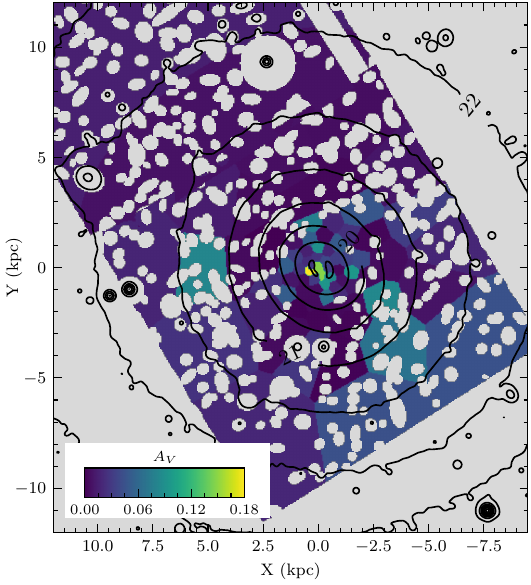}\hfill
\includegraphics[width=0.99\linewidth]{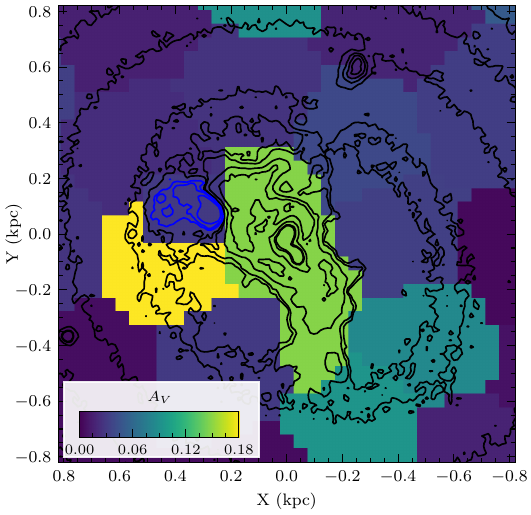}
\caption{\textit{Top panel: }Stellar extinction map of the galaxy.  
\textit{Bottom panel:} Zoom into the central region. 
The blue contours indicate the blue spot location with a metal-rich and intermediate age stellar population (see Fig.\ref{fig:age_zoom-in})}  
\label{fig:Avmap}
\end{figure}

Figure \ref{fig:Avmap} shows the 2D map of the stellar extinction ($A_{V}$) obtained in our analysis (top panel), and a zoom into the innermost $1.6\arcsec \times 1.6 \arcsec$, with the HST contours overplotted (bottom panel). This figure confirms that the extinction only plays a non-negligible role  in the innermost region for at most three Voronoi bins.

\end{document}